\documentclass[lettersize,journal]{IEEEtran}
\IEEEpubidadjcol
\usepackage{xcolor}
\usepackage{amsmath,amsfonts}
\usepackage{amsthm}
\usepackage{algorithmic}
\usepackage{algorithm}
\usepackage{array}
\usepackage[caption=false,font=scriptsize,labelfont=rm,textfont=rm]{subfig}
\usepackage{textcomp}
\usepackage{stfloats}
\usepackage{url}
\usepackage{verbatim}
\usepackage{graphicx}
\usepackage{cite}
\usepackage{amssymb}
\usepackage[colorlinks=true,linkcolor=blue,citecolor=blue,urlcolor=blue]{hyperref}

\theoremstyle{definition}
\newtheorem{remark}{Remark}
\newtheorem{proposition}{Proposition}
\newtheorem{corollary}{Corollary}

\newcommand{\pll}{\kern 0.56em/\kern -0.8em /\kern 0.56em}

\begin{document}

\title{Fisher Information Characterization in GNSS DPE under Multipath-Induced Channel Uncertainty}

\author{Sijia Li,~\IEEEmembership{Graduate Student Member,~IEEE}, Bing Xu,~\IEEEmembership{Senior Member,~IEEE}, Shuo Tang,\\ and Pau Closas,~\IEEEmembership{Senior Member,~IEEE}

\thanks{The work described in this paper was supported by a grant from the Research Grants Council of the Hong Kong Special Administrative Region, China (PolyU15203326), and the National Science Foundation under Awards 1845833, 2326559, and 2530870.}
\thanks{S. Li and B. Xu are with the Department of Aeronautical and Aviation Engineering, The Hong Kong Polytechnic University, Hong Kong. (e-mail: sijia-franz.li@connect.polyu.hk, pbing.xu@polyu.edu.hk).}
\thanks{S. Tang and P. Closas are with the Department of Electrical and Computer Engineering, Northeastern University, Boston, MA 02115, USA. (e-mail: \{tang.shu, closas\}@northeastern.edu).}}

\maketitle

\begin{abstract}
The availability and utilization of Fisher information in global navigation satellite system (GNSS) direct position estimation (DPE) under multipath-induced channel uncertainty are investigated. First, a unified estimation-theoretic framework is developed to characterize information mapping and utilization across direct and two-step receivers. Next, a scalar-equivalent statistical channel model for short-delay multipath is established via least-squares (LS) projection. Based on this model, receiver-state Cram\'er-Rao bounds (CRBs) are derived under perfect and statistical channel knowledge, while misspecified CRBs (MCRBs) are derived under a simplified deterministic complex-gain model that neglects channel uncertainty. The closed-form approximations and asymptotic analyses of the derived bounds further reveal how channel uncertainty affects different components of waveform-domain Fisher information. Our results reveal that: 1) DPE and fully coupled two-step estimation achieve comparable accuracy, whereas the performance loss caused by sequential decoupling substantially exceeds the additional loss from suboptimal weighting; 2) under statistical channel knowledge, improving received signal quality does not necessarily yield noticeable positioning gains; 3) positioning information remains available even as the coherent channel mean vanishes; and 4) neglecting channel randomness causes the MCRBs to diverge in this limit and can introduce positioning error floors that become more pronounced with longer observations.
\end{abstract}


\section{Introduction}\label{section_1}
Global navigation satellite system (GNSS) has been pivotal in providing positioning, navigation, and timing (PNT) services across a broad range of applications~\cite{sec1-ref1,sec1-ref2}. The growing reliance on such services has placed more stringent requirements on the availability, continuity, and robustness of GNSS receivers, especially in densely populated urban environments where complex navigation conditions and dynamic obstacles pose substantial challenges~\cite{sec1-ref3}. In this context, direct position estimation (DPE) has emerged as a promising receiver architecture for robust GNSS positioning, representing a one-step alternative to the conventional two-step positioning framework~\cite{sec1-ref4}. Unlike two-step receivers, which first estimate the synchronization parameters such as code delays and Doppler shifts associated with individual satellite vehicles (SVs) and subsequently infer the navigation state from these intermediate measurements, DPE directly maximizes the joint likelihood of the received signals over the navigation state, thereby allowing the multi-SV signal contributions to be collectively exploited under a common receiver-state constraint~\cite{sec1-ref5,sec1-ref6}.

Nevertheless, the performance advantage of DPE over conventional two-step receivers does not simply arise from its one-step implementation~\cite{sec1-ref7,sec1-ref8,sec1-ref9,sec1-ref10}. It has been proved that if the first-stage estimates are asymptotically efficient and the second-stage estimation is performed with Fisher-optimal weighting while preserving the full dependency of the synchronization parameters on the navigation state, the resulting fully coupled two-step estimator is asymptotically equivalent to DPE~\cite{sec1-ref9,sec1-ref10}. However, conventional two-step GNSS receivers adopt a sequentially decoupled estimation architecture, in which the position-related state is first estimated solely from the code delay measurements, followed by the estimation of the velocity-related state from the Doppler shift measurements conditioned on the preceding position estimate. Such structural decoupling prevents the Doppler estimates from directly contributing to the position-related state update and therefore discards the Doppler-induced position information~\cite{sec1-ref11,sec1-ref12,sec1-ref13}.

Despite this structural advantage in information utilization, DPE remains susceptible to multipath propagation, particularly in urban canyons where closely spaced reflections are difficult to resolve due to the limited bandwidth of GNSS signals~\cite{sec1-ref21}. Under such short-delay multipath conditions, the superposition of direct and reflected signal components introduces amplitude and carrier-phase variations as well as waveform distortion~\cite{sec1-ref22,sec1-ref23}. When the receiver cannot accurately characterize the instantaneous multipath channel response, these effects collectively give rise to channel uncertainty that can potentially degrade positioning accuracy. Existing studies have mainly addressed this issue through receiver-side algorithmic designs that either suppress or exclude multipath-contaminated signal contributions in the navigation domain~\cite{sec1-ref14,sec1-ref15,sec1-ref16,sec1-ref17}, as well as through the incorporation of prior state information from auxiliary sensors such as inertial navigation systems (INS)~\cite{sec1-ref18}. While these approaches can effectively mitigate multipath-induced positioning errors under specific propagation conditions, the resulting improvements in positioning accuracy do not directly characterize the fundamental performance limits under multipath propagation and the associated channel uncertainty. Early theoretical studies on GNSS DPE have primarily focused on characterizing the fundamental limits of estimation accuracy through the Cram\'er-Rao bound (CRB) established in~\cite{sec1-ref5} and the Ziv-Zakai bound (ZZB) analysis in~\cite{sec1-ref6}, both derived under idealized signal models without explicitly accounting for multipath-induced channel uncertainty. Consequently, the impact of multipath-induced channel uncertainty on the Fisher information exploitable for GNSS DPE under different levels of channel knowledge remains insufficiently characterized.

Against the above background, this paper develops an information-centric analysis of GNSS DPE under channel uncertainty arising from short-delay multipath propagation. In particular, we investigate how different levels of channel knowledge shape the Fisher information carried by the received waveform, how this information is mapped into receiver state information through satellite geometry, and how effectively the resulting estimation information can be exploited by different receiver architectures. Motivated by the statistical variability of multipath propagation in complex urban environments~\cite{sec1-ref24}, we adopt a Gaussian wide-sense stationary uncorrelated scattering (WSSUS) channel model to characterize the aggregate channel effects~\cite{sec1-ref20}. The main contributions of this paper are summarized as follows:
\begin{itemize}
    \item We establish a scalar-equivalent statistical channel model for short-delay multipath via least-squares (LS) projection. Under the Gaussian WSSUS assumption, we derive the second-order statistics of the equivalent complex channel gain and numerically evaluate the validity of neglecting the orthogonal waveform-distortion residual.
    \item  We establish a unified estimation-theoretic framework to characterize Fisher information utilization across different GNSS receiver architectures. We demonstrate the asymptotic equivalence between DPE and the fully coupled two-step estimator under efficient first-stage estimation and Fisher-optimal weighting. Moreover, we illustrate that the sequentially decoupled architecture incurs a positive semidefinite increase in the asymptotic error covariance.
    \item  We derive receiver-state CRBs under correctly specified channel models with perfect and statistical channel knowledge. Closed-form expressions reveal how channel uncertainty affects different waveform-information components and identify the information exploitable under limited channel knowledge. Asymptotic analysis of the derived CRBs further characterizes positioning performance across different levels of channel randomness.
    \item We derive receiver-state misspecified CRBs (MCRBs) for receivers that neglect channel uncertainty and assume an unknown deterministic complex gain for each SV. Closed-form expressions characterize the information loss due to unknown channel gains and the additional estimation error induced by unmodeled random channel variations.
    \item Numerical results validate the derived performance bounds and reveal the following insights: 1) DPE and fully coupled two-step estimation achieve nearly identical accuracy, whereas sequential decoupling incurs increasing performance loss with observation duration and substantially exceeds the additional loss from suboptimal weighting due to unutilized Doppler-related position information; 2) under statistical channel knowledge, improved signal quality may yield limited positioning gains, and positioning information remains available through the observation covariance even as the coherent mean vanishes; 3) neglecting channel randomness prevents the exploitation of such information, causing the MCRB to diverge in this limit; and 4) model misspecification introduces positioning error floors that persist despite improved signal quality and become more pronounced with longer observations.
\end{itemize}

The remainder of this paper is organized as follows. Section~\ref{section_2} develops the signal and channel model. Section~\ref{section_3} characterizes Fisher information mapping and utilization across different receiver architectures. Section~\ref{section_4} derives the CRBs under correctly specified channel models with different levels of channel knowledge, whereas Section~\ref{section_5} derives the MCRBs under channel model misspecification. Section~\ref{section_6} discusses the numerical results, and Section~\ref{section_7} concludes the paper.

\emph{Notations:} Scalars are denoted in standard typeface, while vectors and matrices are represented by lowercase boldface and uppercase boldface, respectively. The sets of complex and real numbers are denoted by $\mathbb{C}$ and $\mathbb{R}$, respectively. The inverse, transpose, conjugate, conjugate transpose, trace, and determinant operations are represented by $\left(\cdot\right)^{-1}$, $\left(\cdot\right)^T$, $\left(\cdot\right)^*$, $\left(\cdot\right)^H$, $\mathrm{tr}\left(\cdot\right)$, and $\det\left(\cdot\right)$, respectively. The absolute value, Euclidean norm, and weighted norm are written as $\vert\cdot\vert$, $\Vert\cdot\Vert$, and $\Vert\cdot\Vert_{\mathbf{W}}$, respectively. The real part and imaginary part of a complex number are denoted by $\Re\left\{\cdot\right\}$ and $\Im\left\{\cdot\right\}$, respectively. The Hadamard product and the Kronecker product are denoted by $\odot$ and $\otimes$, respectively. An $N \times N$ identity matrix is denoted by $\mathbf{I}_N$. A diagonal matrix with the same value as the vector $\mathbf{x}$ on the diagonal is represented as $\mathrm{diag}\left(\mathbf{x}\right)$. The notation $\mathcal{CN}\left(\mu,\sigma^2\right)$ denotes a circularly symmetric complex Gaussian distribution with mean $\mu$ and variance $\sigma^2$, and $\nabla_v(\cdot)$ denotes the derivative of a matrix or scalar function with respect to the variable $v$. Finally, $\delta(\cdot)$ is the Dirac delta pseudo-function, and $\mathbb{E}[\cdot]$ denotes the e statistic expectation of a random variable.

\section{Signal and Channel Model}\label{section_2}
Consider a scenario where $K$ SVs contribute to the positioning, with their indices collected in $\mathcal{K} = \left\{1,\ldots,K\right\}$. For the $k$-th SV, its position, velocity, and clock bias are denoted by $\mathbf{p}_k = \left[p_{x,k},p_{y,k},p_{z,k}\right]^T$, $\mathbf{v}_k = \left[v_{x,k},v_{y,k},v_{z,k}\right]^T$, and $\delta t_k$ respectively, which are known to the receiver via ephemeris and navigation message. The target is positioned at an unknown location $\mathbf{p} = \left[p_{x},p_{y},p_{z}\right]^T$ with an unknown velocity $\mathbf{v} =\left[v_{x},v_{y},v_{z}\right]^T$, both of which are to be estimated. Moreover, the receiver is subject to an unknown clock bias $\delta t$, which is included as part of the navigation-domain receiver state $\boldsymbol{\xi}=\left[\mathbf{p}^T,\mathbf{v}^T,\delta t\right]^T$ along with its position and velocity. 

At the transmitter side, each SV transmits low-rate navigation data, which is spread by the corresponding pseudo-random noise (PRN) code using direct-sequence spread spectrum (DS-SS) for reliable reception under low signal-to-noise ratio (SNR) conditions. Let $M$ denote the number of chips within a code and $R_\mathrm{chip} = 1/T_\mathrm{chip}$ denote the chip rate. Therefore, the code duration is given by $T_\mathrm{code} = M/R_\mathrm{chip}$. Consider the reception over $P$ consecutive code periods during which the navigation data remain constant, the baseband transmit signal from the $k$-th SV within the single code period is given by
\begin{equation}\label{sec2-eq1}
     x_{k}(t) = d_k\sum_{m=0}^{M-1}c_{m,k} g\left(t-mT_\mathrm{chip}\right),
\end{equation}
for $t\in[0,T_\mathrm{code})$, where $d_k\in\left\{\pm1\right\}$ denotes the navigation bit contained in the signal transmitted by the $k$-th SV, $c_{m,k}$ denotes the $m$-th chip of the unique PRN code transmitted by the $k$-th SV, and $g(t)$ denotes the real-valued  impulse response of the transmission analog filter. Therefore, the noiseless line-of-sight (LoS) signal from the $k$-th SV observed at the positioning target can be expressed as
\begin{equation}\label{sec2-eq2}
    s_k(t) = \sum_{p=0}^{P-1}x_{k}\left(t-pT_\mathrm{code}-\tau_{k}\right)e^{\jmath2\pi(\nu_{k}t-f_c\tau_k)},
\end{equation}
where $f_c$ denotes the carrier frequency of the GNSS signal. The deterministic delay $\tau_k$ and Doppler shift $\nu_k$ associated with the $k$-th SV are determined by the relative geometry and motion between the SV and the positioning target, and are respectively given by
\begin{align}
     \tau_{k} &= \frac{1}{c} \Vert \mathbf{p}_{k} - \mathbf{p} \Vert + \delta t - \delta t_k + \epsilon_k, \label{sec2-eq3}\\
     \nu_{k} &= -\left(\mathbf{v}_k-\mathbf{v}\right)^T \frac{\mathbf{p}_k-\mathbf{p}}{\Vert\mathbf{p}_k-\mathbf{p}\Vert} \frac{f_c}{c},\label{sec2-eq4}
\end{align}
 where $c$ denotes the speed of light and $\epsilon_k$ denotes the overall timing error, including the atmospheric delay and ephemeris error of the $k$-th SV. Such errors can be substantially mitigated by standard GNSS correction techniques, and are assumed negligible in the subsequent analysis. Therefore, the $p$-th code period of the GNSS signal transmitted by the $k$-th SV at time $t = t+pT_\mathrm{code}$ can be expressed as
 \begin{equation}\label{sec2-eq5}
    s_{p,k}(t) = x_k(t-\tau_k)e^{\jmath2\pi(\nu_k(t+pT_\mathrm{code})-f_c\tau_k)}.
\end{equation}

\begin{figure}[t]
    \centering
    \includegraphics[width=0.7078069\linewidth]{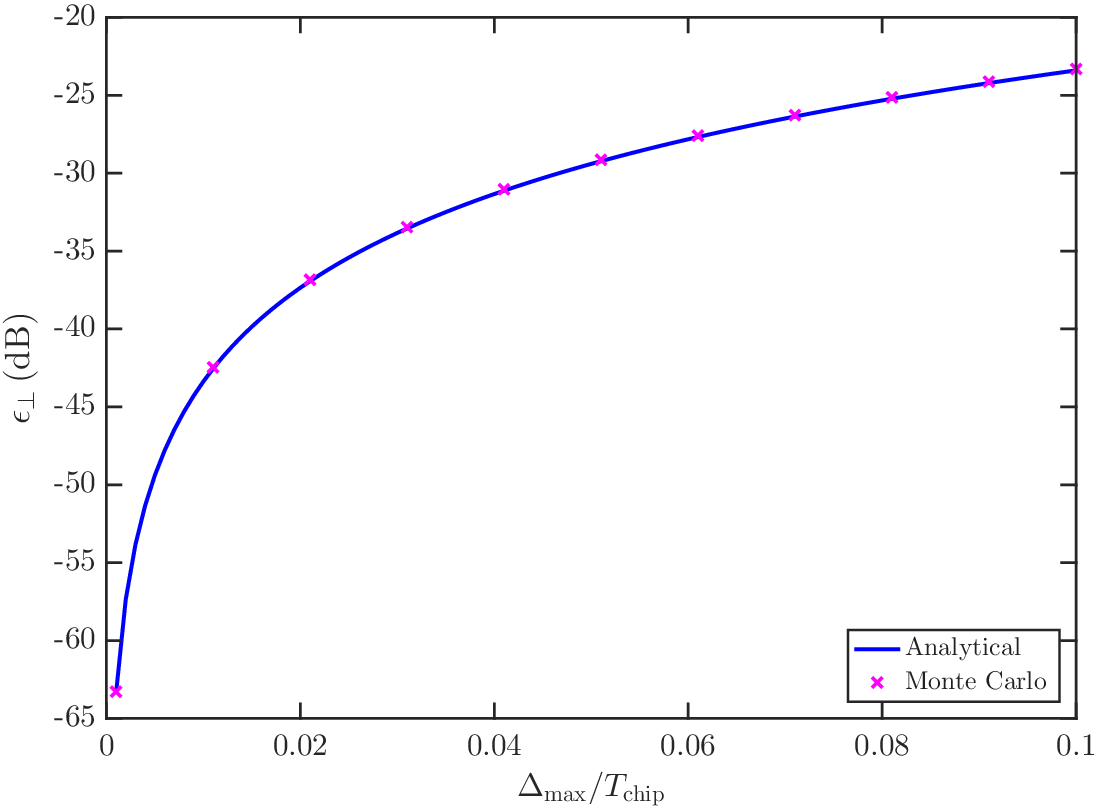}
    \caption{Normalized orthogonal projection residual energy versus maximum multipath excess delay.}
    \label{figure_8}
\end{figure}

In practice, the propagation channel associated with the $k$-th SV is modeled as a time-varying stochastic multipath channel with an $L_k$-path impulse response given by
\begin{equation}\label{sec2-eq6}
    h_k(t,\Delta\tau) = \sum_{l=0}^{L_k-1}\alpha_{l,k}e^{\jmath\phi_{l,k}}e^{\jmath2\pi\Delta\nu_{l,k}t}\delta(\Delta\tau-\Delta\tau_{l,k}),
\end{equation}
where $\alpha_{l,k}$ and $\phi_{l,k}$ denote the real-valued gain and initial phase of the $l$-th propagation path associated with the $k$-th SV, respectively, and $\Delta\nu_{l,k}$ and $\Delta\tau_{l,k}$ denote its Doppler offset and excess delay relative to the LoS component. Assuming that the channel remains quasi-static over each code period, the channel impulse response during the $p$-th code period can be approximated as
\begin{equation}\label{sec2-eq7}
    \begin{aligned}
        h_{p,k}(\Delta\tau)&\triangleq h_{k}(pT_{\mathrm{code}},\Delta\tau)=
        \underbrace{
        \alpha_{0,k}e^{\jmath\phi_{0,k}}\delta(\Delta\tau)}_{\mu_{h,k}(\Delta\tau)}
        \\&\!\!\!\!\!\!\!\!\!+\underbrace{
        \sum_{l=1}^{L_k-1}
        \alpha_{l,k}e^{\jmath(\phi_{l,k}+2\pi\Delta\nu_{l,k}pT_\mathrm{code})}\delta(\Delta\tau - \Delta\tau_{l,k})}_{w_{p,k}(\Delta\tau)},
    \end{aligned}
\end{equation}
where $\mu_{h,k}(\Delta\tau)$ denotes the deterministic coherent component and $w_{p,k}(\Delta\tau)$ represents the zero-mean diffuse-scattering component\footnote{The zero-mean property of $w_{p,k}(\Delta\tau)$ follows from the assumption that the initial phase of each scattering path is uniformly distributed over $[0,2\pi)$, which leads to $\mathbb{E}[e^{\jmath\phi_{l,k}}] = 0$. This assumption is justified and commonly adopted in rich multipath environments~\cite{sec2-ref5,sec2-ref3,sec2-ref4}.}. Under the WSSUS assumption, the second-order statistic of $w_{p,k}(\Delta\tau)$ is characterized by~\cite{sec2-ref1,sec2-ref2}
\begin{equation}\label{sec2-eq8}
    \mathbb{E}[w_{p,k}(\Delta\tau)w_{p,k}^*(\Delta\tau')] = \Sigma_{h,k}\varphi_k(\Delta\tau)\delta(\Delta\tau-\Delta\tau'),
\end{equation}
where $\varphi_k(\Delta\tau)$ denotes the normalized power delay profile (PDP) and $\Sigma_{h,k}$ denotes the total power of the diffuse-scattering component. The corresponding noiseless received signal is then obtained by convolving the LoS waveform with the multipath channel impulse response, which yields
\begin{equation}\label{sec2-eq9}
    \begin{aligned}
        &r_{p,k}(t) = \beta_{0,k}s_{p,k}(t) + \sum_{l=1}^{L_k-1}\beta_{l,p,k}s_{p,k}(t-\Delta\tau_{l,k})\\
        &\approx\left(
        \beta_{0,k}+\!\!\sum_{l=1}^{L_k-1}\beta_{l,p,k}
        \right)s_{p,k}(t) - \left(
        \sum_{l=1}^{L_k-1}\beta_{l,p,k}\Delta\tau_{l,k}
        \right)\dot{s}_{p,k}(t),
    \end{aligned}
\end{equation}
where $\beta_{0,k}\triangleq\alpha_{0,k}e^{\jmath\phi_{0,k}}$,
$\beta_{l,p,k}\triangleq\alpha_{l,k}e^{\jmath(\phi_{l,k}+2\pi\Delta\nu_{l,k}pT_\mathrm{code})}$, and $\dot{s}_{p,k}(t)$ denotes the time derivative of the LoS waveform, which can be expressed as
\begin{equation}\label{sec2-eq20}
    \dot{s}_{p,k}(t) = e^{\jmath\psi_{p,k}(t)}\dot{x}_k(t-\tau_k)+\jmath2\pi\nu_ks_{p,k}(t),
\end{equation}
where $\psi_{p,k}(t)\triangleq2\pi(\nu_k(t+pT_\mathrm{code})-f_c\tau_k)$. The approximation in~\eqref{sec2-eq9} follows from the first-order Taylor expansion of the delayed waveform considering $\Delta\tau_{l,k}$ is sufficiently small. Substituting~\eqref{sec2-eq20} into~\eqref{sec2-eq9} yields 
\begin{equation}\label{sec2-eq21}
    \begin{aligned}
        r_{p,k}(t) &\approx\underbrace{\left(\beta_{0,k}+\sum_{l = 1}^{L_k-1}\beta_{l,p,k}-\jmath2\pi\nu_k\sum_{l=1}^{L_k-1}\beta_{l,p,k}\Delta\tau_{l,k}\right)s_{p,k}(t)}_{\text{Complex-scaled LoS component}} \\&
        - \underbrace{\left(\sum_{l=1}^{L_k-1}\beta_{l,p,k}\Delta\tau_{l,k}\right)e^{\jmath\psi_{p,k}(t)}\dot{x}_k(t-\tau_k)}_{\text{Orthogonal waveform distortion}}.
    \end{aligned}
\end{equation}
It follows from~\eqref{sec2-eq21} that under the short-delay multipath assumption, the received signal can be approximated as the sum of a complex-scaled replica of the LoS waveform and a waveform distortion proportional to the derivative of the baseband signal, which is orthogonal to the LoS waveform due to the real-valued and periodic signal properties. This decomposition naturally motivates obtaining the scalar-equivalent complex channel gain through the LS projection of the noiseless received waveform onto the LoS waveform, yielding
\begin{equation}\label{sec2-eq10}
    \begin{aligned}
        \tilde{h}_{p,k} &= \arg\min_{a\in\mathbb{C}}\int_{\mathcal{T}_p}\vert
        r_{p,k}(t)-as_{p,k}(t)
        \vert^2\,dt\\
        &=\int h_{p,k}(\Delta\tau)\mathcal{A}_{k}(\Delta\tau)\,d\Delta\tau,
    \end{aligned}
\end{equation}
where $\mathcal{A}_{k}(\Delta\tau)\triangleq\frac{1}{E_s}\int_{\mathcal{T}_p}s_{p,k}(t-\Delta\tau)s^*_{p,k}(t)\,dt$ denotes the normalized autocorrelation function (ACF) of the LoS waveform transmitted by the $k$-th SV, $E_s = \int_{\mathcal{T}_p}\vert s_{p,k}(t)\vert^2\,dt$ denotes the signal energy over a single code period, and $\mathcal{T}_p$ denotes the corresponding observation interval. Using~\eqref{sec2-eq8}, the mean and covariance of $\tilde{h}_{p,k}$ can be derived as
\begin{align}
    &\mu_{\tilde{h},k} = \beta_{0,k}\mathcal{A}_k(0),\label{sec2-eq18}\\
    &\sigma^2_{\tilde{h},k} = \Sigma_{h,k}\int \varphi_k(\Delta\tau) \vert \mathcal{A}_k(\Delta\tau)\vert^2 \,d\Delta\tau.\label{sec2-eq19}
\end{align}
Based on~\eqref{sec2-eq10}, the noiseless received waveform admits the following orthogonal decomposition
\begin{equation}\label{sec2-eq11}
    r_{p,k}(t) = \tilde{h}_{p,k}s_{p,k}(t) + e_{p,k}^\perp(t),
\end{equation}
where $e_{p,k}^\perp(t)$ denotes the residual waveform distortion that satisfies $\int_{\mathcal{T}_p}e_{p,k}^\perp(t)s^*_{p,k}(t)\,dt = 0$. Fig.~\ref{figure_8} illustrates the normalized residual energy $\epsilon_\perp$ as a function of the maximum multipath excess delay $\Delta_{\max}$, where
\begin{equation}\label{sec2-eq12}
    \epsilon_\perp\triangleq\frac{\mathbb{E}\left[\int_{\mathcal{T}_p}|e_{p,k}^\perp(t)|^2\,dt\right]}{\mathbb{E}\left[\int_{\mathcal{T}_p}|r_{p,k}(t)|^2\,dt\right]}.
\end{equation}
For instance, for legacy GPS L1 C/A signals, it can be observed that the residual energy remains below approximately $-23$ dB when $\Delta_\mathrm{max}\leq0.1 \,T_{\mathrm{chip}}$, which corresponds to a maximum excess path length of $29.3$ m, and accounts for less than $0.5\%$ of the total received signal energy. Therefore, under the considered short-delay multipath conditions, we neglect the orthogonal residual and approximate the noiseless received waveform as 
\begin{equation}\label{sec2-eq13}
    r_{p,k}(t)\approx\tilde{h}_{p,k}s_{p,k}(t).
\end{equation}
At the receiver side, sampling the received signal at $T_{\mathrm{samp}}$ yields the discrete-time signal model, which is given by
\begin{equation}\label{sec2-eq14}
    \mathbf{y}_{p,k} = \tilde{h}_{p,k}\mathbf{s}_{p,k}+\mathbf{z}_{p,k},
\end{equation}
where $\mathbf{z}_{p,k} \sim \mathcal{CN}\big(\mathbf{0},\sigma_{z,k}^2\mathbf{I}_{N}\big)$ denotes the additive white Gaussian noise (AWGN) vector associated with the reception of the $k$-th SV, with $N = T_{\mathrm{code}}/T_{\mathrm{samp}}$ representing the number of sampling points. $\mathbf{s}_{p,k} \in \mathbb{C}^{N\times1}$ denotes the noiseless LoS waveform of the $p$-th code transmitted by the $k$-th SV, which can be decomposed as
\begin{equation}\label{sec2-eq15}
    \mathbf{s}_{p,k} = e^{-\jmath2\pi f_c\tau_k}
    \left(
    \mathbf{x}_k\odot e^{\jmath2\pi\nu_k\mathbf{t}_p}
    \right),
\end{equation}
where $\mathbf{t}_{p} \triangleq [0,\ldots,(N-1)T_{\mathrm{samp}}]^T + pT_\mathrm{code}\mathbf{1}_N$ collects the sampling times in the $p$-th code period and $\mathbf{x}_k$ denotes the delayed envelope.
Let $\mathbf{y}_k = \big[\mathbf{y}_{0,k}^T,\ldots,\mathbf{y}_{P-1,k}^T\big]^T$ denote the vector collecting overall signals received from the $k$-th SV, which is given by
\begin{equation}\label{sec2-eq16}
    \mathbf{y}_k = \mathbf{S}_k\tilde{\mathbf{h}}_k+\mathbf{z}_k,
\end{equation}
where $\mathbf{S}_k\triangleq\mathrm{blkdiag}(\mathbf{s}_{0,k},\ldots,\mathbf{s}_{P-1,k})\in\mathbb{C}^{PN\times P}$ denotes the block-diagonal waveform matrix. The equivalent channel vector $\tilde{\mathbf{h}}_k\in\mathbb{C}^{P\times1}$ is modeled as $\tilde{\mathbf{h}}_k\sim\mathcal{CN}\big(\mu_{\tilde{h},k}\mathbf{1}_P,\sigma_{\tilde{h},k}^2\mathbf{I}_P\big)$, where the channel coefficients are assumed to be independent and identically distributed across the $P$ code periods. Moreover, $\mathbf{z}_k=\big[\mathbf{z}_{0,k}^T,\ldots,\mathbf{z}_{P-1,k}^T\big]^T\sim\mathcal{CN}\big(\mathbf{0},\sigma_{z,k}^2\mathbf{I}_{PN}\big)$ denotes the stacked AWGN vector. Then, noting that $\tilde{\mathbf{h}}_k = \mu_{\tilde{h},k}\mathbf{1}_P + \Delta\tilde{\mathbf{h}}_k$,~\eqref{sec2-eq16} can be rewritten as
\begin{equation}\label{sec2-eq17}
    \mathbf{y}_k = \mu_{\tilde{h},k}\mathbf{s}_k + \mathbf{S}_k\Delta\tilde{\mathbf{h}}_k
    + \mathbf{z}_k,
\end{equation}
where $\mathbf{s}_k = \big[\mathbf{s}_{0,k}^T,\ldots,\mathbf{s}_{P-1,k}^T\big]^T$ and
$\Delta\tilde{\mathbf{h}}_k\sim\mathcal{CN}(\mathbf{0},\sigma_{\tilde{h},k}^2\mathbf{I}_P)$. 
\begin{remark}\label{remark_1}
\textit{(Signal-Space Selectivity of Channel Uncertainty)}
Equation~\eqref{sec2-eq17} shows that the received observations consist of contributions from both the coherent and random channel components, where the latter is confined to the column space $\mathcal{S}_k\triangleq\mathrm{col}(\mathbf{S}_k)=\big\{\big[a_0\mathbf{s}_{0,k}^T,\ldots,a_{P-1}\mathbf{s}_{P-1,k}^T\big]^T:a_p\in\mathbb{C}\big\}$. Consequently, the channel uncertainty arising from the random channel coefficients is confined to $\mathcal{S}_k$, while its orthogonal complement $\mathcal{S}_k^\perp$ remains unaffected by this uncertainty. 
\end{remark}

\section{Information Mapping and Utilization in Direct and Two-step Estimation}\label{section_3}

\subsection{Geometric Mapping of Fisher Information}
To facilitate the subsequent analysis, we first partition the receiver state as $\boldsymbol{\xi}=\big[\boldsymbol{\alpha}^T,\boldsymbol{\beta}^T\big]^T \in \mathbb{R}^{7\times1}$, where $\boldsymbol{\alpha}=\big[\mathbf{p}^T,\delta t\big]^T$ denotes the position-related state, and $\boldsymbol{\beta}=\mathbf{v}$ denotes the velocity-related state. Then, define $\boldsymbol{\zeta}= \big[\boldsymbol{\tau}^T,\boldsymbol{\nu}^T\big]^T \in\mathbb{R}^{2K\times1}$ as the synchronization-parameter vector collecting the delays and Doppler shifts of all participating SVs. Based on~\eqref{sec2-eq3} and~\eqref{sec2-eq4}, the implicit dependency of the synchronization-parameter vector on the receiver state can be expressed as $\boldsymbol{\zeta}\triangleq\boldsymbol{\zeta}(\boldsymbol{\xi}) = 
    \big[
    \boldsymbol{\tau}^T(\boldsymbol{\alpha}),
    \boldsymbol{\nu}^T(\boldsymbol{\alpha},\boldsymbol{\beta})
    \big]^T$. Therefore, the Jacobian matrix of $\boldsymbol{\zeta}$ with respect to $\boldsymbol{\xi}$ is given by
    
\begin{equation}\label{sec3-eq1}
    \mathbf{G}\triangleq \frac{\partial\boldsymbol{\zeta}(\boldsymbol{\xi})}{\partial\boldsymbol{\xi}^T} = 
    \begin{bmatrix}
        \mathbf{G}_{\boldsymbol{\tau}\boldsymbol{\alpha}} & \mathbf{0}\\
        \mathbf{G}_{\boldsymbol{\nu}\boldsymbol{\alpha}} & \mathbf{G}_{\boldsymbol{\nu}\boldsymbol{\beta}}
    \end{bmatrix},
\end{equation}
where $\mathbf{G}_{\boldsymbol{\tau}\boldsymbol{\alpha}} = \frac{\partial \boldsymbol{\tau}(\boldsymbol{\alpha})} {\partial \boldsymbol{\alpha}^T}$, $\mathbf{G}_{\boldsymbol{\nu}\boldsymbol{\alpha}} = \frac{\partial \boldsymbol{\nu}(\boldsymbol{\alpha},\boldsymbol{\beta})}{\partial \boldsymbol{\alpha}^T}$, and $\mathbf{G}_{\boldsymbol{\nu}\boldsymbol{\beta}} = \frac{\partial \boldsymbol{\nu}(\boldsymbol{\alpha},\boldsymbol{\beta})}{\partial \boldsymbol{\beta}^T}$. The zero block in~\eqref{sec3-eq1} follows from the independence of the code delays from the receiver velocity, whereas $\mathbf{G}_{\boldsymbol{\nu}\boldsymbol{\alpha}}$ captures the dependence of the Doppler shifts on the position-related state through the satellite geometry.

Let $\boldsymbol{y} = \big[\mathbf{y}_{1}^T,\ldots,\mathbf{y}_{K}^T\big]^T$ denote the stacked received observations from all SVs, and let $\mathcal{L}(\boldsymbol{y}\,|\,\boldsymbol{\zeta})$ denote the log-likelihood function of $\boldsymbol{y}$ conditioned on $\boldsymbol{\zeta}$. The Fisher information matrix (FIM) with respect
to $\boldsymbol{\zeta}$ is then defined as $\mathbf{F}(\boldsymbol{\zeta})\triangleq\mathbb{E}_{\boldsymbol{y}\,|\,\boldsymbol{\zeta}}\Big[
        \nabla_{\boldsymbol{\zeta}}\mathcal{L}(\boldsymbol{y}\,|\,\boldsymbol{\zeta})
        \nabla_{\boldsymbol{\zeta}}^T\mathcal{L}(\boldsymbol{y}\,|\,\boldsymbol{\zeta})
        \Big]$. 
Accordingly, $\mathbf{F}(\boldsymbol{\zeta})$ provides an overall characterization of the information in the received observations that is locally exploitable for estimating the synchronization parameters. Since the probability distribution of the observation depends on the receiver state solely through $\boldsymbol{\zeta}(\boldsymbol{\xi})$, the corresponding score functions satisfy $\nabla_{\boldsymbol{\xi}}\mathcal{L}(\boldsymbol{y}\,|\,\boldsymbol{\xi}) = \mathbf{G}^T\nabla_{\boldsymbol{\zeta}}\mathcal{L}(\boldsymbol{y}\,|\,\boldsymbol{\zeta})$. Therefore, the FIM with respect to $\boldsymbol{\xi}$ can be expressed as
\begin{equation}
    \begin{aligned}\label{sec3-eq2}
        &\mathbf{F}(\boldsymbol{\xi}) \triangleq 
        \mathbb{E}_{\boldsymbol{y}\,|\,\boldsymbol{\xi}}\Big[
        \nabla_{\boldsymbol{\xi}}\mathcal{L}(\boldsymbol{y}\,|\,\boldsymbol{\xi})
        \nabla_{\boldsymbol{\xi}}^T\mathcal{L}(\boldsymbol{y}\,|\,\boldsymbol{\xi})
        \Big]\\
        &=\mathbf{G}^T 
        \mathbb{E}_{\boldsymbol{y}\,|\,\boldsymbol{\zeta}}\Big[
        \nabla_{\boldsymbol{\zeta}}\mathcal{L}(\boldsymbol{y}\,|\,\boldsymbol{\zeta})
        \nabla_{\boldsymbol{\zeta}}^T\mathcal{L}(\boldsymbol{y}\,|\,\boldsymbol{\zeta})
        \Big]
        \mathbf{G}=\mathbf{G}^T\mathbf{F}({\boldsymbol{\zeta}})\mathbf{G}.
    \end{aligned}
\end{equation}

\subsection{Information Utilization across Receiver Architectures}
Let $\hat{\boldsymbol{\xi}}_{\mathrm{DPE}}$ denote the $P$-snapshot DPE estimator. The objective of DPE is to directly estimate the receiver state from the received observations under the maximum likelihood (ML) criterion, i.e., 
\begin{equation}\label{sec3-eq3}
    \hat{\boldsymbol{\xi}}_{\mathrm{DPE}} = \arg\max_{\boldsymbol{\xi}}\mathcal{L}(\boldsymbol{y}\,|\,\boldsymbol{\xi}).
\end{equation}
Under standard regularity and full-rank conditions, the DPE estimator is asymptotically efficient and therefore attains the inverse receiver-state FIM, yielding
\begin{equation}\label{sec3-eq4}
    \lim_{P\rightarrow\infty}\mathbf{\Sigma}(\hat{\boldsymbol{\xi}}_{\mathrm{DPE}})
    = \mathbf{F}^{-1}(\boldsymbol{\xi})=\left(
    \mathbf{G}^T\mathbf{F}({\boldsymbol{\zeta}})\mathbf{G}
    \right)^{-1}.
\end{equation}
In contrast, the two-step approach first obtains the ML estimates of the intermediate synchronization parameters, i.e.,
\begin{equation}\label{sec3-eq5}
    \hat{\boldsymbol{\zeta}}  = \arg\max_{\boldsymbol{\zeta}}
    \mathcal{L}(\boldsymbol{y}\,|\,\boldsymbol{\zeta}).
\end{equation}
Similarly, the first-stage ML estimator in~\eqref{sec3-eq5} is asymptotically efficient, such that 
\begin{equation}\label{sec3-eq6}
    \lim_{P\rightarrow\infty}\mathbf{\Sigma}(\hat{\boldsymbol{\zeta}}) = \mathbf{F}^{-1}(\boldsymbol{\zeta}).
\end{equation}

Alternatively, let $\hat{\boldsymbol{\xi}}_{\mathrm{FCE}}$ denote the $P$-snapshot fully coupled two-step estimator where the second-stage estimation jointly preserves the complete dependence of the synchronization parameters on the receiver state. The resulting asymptotically efficient receiver-state estimate can be obtained via the Fisher-weighted least squares (WLS) problem, given as~\cite{sec1-ref10}
\begin{equation}\label{sec3-eq7}
    \hat{\boldsymbol{\xi}}_{\mathrm{FCE}}
    = \arg\min_{\boldsymbol{\xi}} 
    \left\Vert
    \hat{\boldsymbol{\zeta}} - \boldsymbol{\zeta}(\boldsymbol{\xi})
    \right\Vert_{\mathbf{F}(\boldsymbol{\zeta})}^2 ,
\end{equation}
known to be ML-optimal under Gaussian likelihoods.
We then characterize the local error propagation of the corresponding estimator. Let $\bar{\boldsymbol{\xi}} = \big[\bar{\mathbf{p}}^T,\bar{\mathbf{v}}^T,\bar{\delta t}\big]^T$ denote the true value of the receiver state. By reparameterizing $\boldsymbol{\xi} = \bar{\boldsymbol{\xi}} + \Delta\boldsymbol{\xi}$, the above WLS problem can be rewritten as
\begin{equation}\label{sec3-eq8}
    \Delta\hat{\boldsymbol{\xi}}_\mathrm{FCE} = \arg\min_{\Delta\boldsymbol{\xi}}
    \left\Vert
    \hat{\boldsymbol{\zeta}} - \boldsymbol{\zeta}
    \left(
    \bar{\boldsymbol{\xi}} + \Delta\boldsymbol{\xi}
    \right)
    \right\Vert_{\mathbf{F}(\boldsymbol{\zeta})}^2.
\end{equation}
Denote the first-stage synchronization-parameter estimation error by $\Delta\boldsymbol{\zeta} = \hat{\boldsymbol{\zeta}} - \boldsymbol{\zeta}(\bar{\boldsymbol{\xi}})$. Then, the residual vector in~\eqref{sec3-eq8} can be approximated as
\begin{equation}
    \begin{aligned}\label{sec3-eq9}
        \hat{\boldsymbol{\zeta}} - \boldsymbol{\zeta}
        \left(
        \bar{\boldsymbol{\xi}} + \Delta\boldsymbol{\xi}
        \right) &= \boldsymbol{\zeta}(\bar{\boldsymbol{\xi}}) + \Delta\boldsymbol{\zeta} - \boldsymbol{\zeta}
        \left(
        \bar{\boldsymbol{\xi}} + \Delta\boldsymbol{\xi}
        \right)\\
        &\overset{(a)}{\approx} 
        \Delta\boldsymbol{\zeta} - \mathbf{G}\Delta\boldsymbol{\xi},
    \end{aligned}
\end{equation}
where $(a)$ follows from the first-order Taylor expansion of $\boldsymbol{\zeta}
    \left(
    \bar{\boldsymbol{\xi}} + \Delta\boldsymbol{\xi}
    \right)$ around $\bar{\boldsymbol{\xi}}$. Therefore, the local WLS approximation of~\eqref{sec3-eq8} is given by
\begin{equation}\label{sec3-eq10}
    \Delta\hat{\boldsymbol{\xi}}_\mathrm{FCE} =
    \arg\min_{\Delta\boldsymbol{\xi}}
    \left\Vert
     \Delta\boldsymbol{\zeta} - \mathbf{G}\Delta\boldsymbol{\xi}
    \right\Vert_{\mathbf{F}(\boldsymbol{\zeta})}^2.
\end{equation}
Taking the derivative of the cost function in~\eqref{sec3-eq10} with respect to $\Delta\boldsymbol{\boldsymbol{\xi}}$ and setting it to zero yields
\begin{equation}\label{sec3-eq11}
    \Delta\hat{\boldsymbol{\xi}}_{\mathrm{FCE}}
    =
    \underbrace{
    \left(
    \mathbf{G}^T
    \mathbf{F}(\boldsymbol{\zeta})
    \mathbf{G}
    \right)^{-1}
    \mathbf{G}^T
    \mathbf{F}(\boldsymbol{\zeta})
    }_{\mathbf{E}_{\mathrm{opt}}}
    \Delta\boldsymbol{\zeta},
\end{equation}
where $\mathbf{E}_{\mathrm{opt}}$ denotes the Fisher-optimal local error propagation matrix. Therefore, the asymptotic covariance of the corresponding estimator is given by
\begin{equation}\label{sec3-eq12}
    \lim_{P\rightarrow\infty}\mathbf{\Sigma}(\hat{\boldsymbol{\xi}}_{\mathrm{FCE}}) = \mathbf{E}_{\mathrm{opt}}
    \mathbf{F}^{-1}(\boldsymbol{\zeta})\mathbf{E}_{\mathrm{opt}}^T=\mathbf{F}^{-1}(\boldsymbol{\xi}),
\end{equation}
which corroborates the asymptotic equivalence of DPE and the fully coupled two-step receiver architecture.

However, conventional GNSS receivers typically adopt a sequentially decoupled second-stage mapping, where the position-related state is first estimated exclusively from the delay estimates, given as
\begin{equation}\label{sec3-eq13}
    \hat{\boldsymbol{\alpha}} = \arg\min_{\boldsymbol{\alpha}}
    \left\Vert
    \hat{\boldsymbol{\tau}} - \boldsymbol{\tau}(\boldsymbol{\alpha})
    \right\Vert_{\mathbf{W}_{\boldsymbol{\tau}}}^2,
\end{equation}
where $\mathbf{W}_{\boldsymbol{\tau}}$ denotes the delay-domain weighting matrix. Using the resulting estimates $\hat{\boldsymbol{\alpha}}$ in the geometric Doppler relation given in~\eqref{sec2-eq4}, the velocity-related state can be subsequently estimated, and is given by
\begin{equation}\label{sec3-eq14}
    \hat{\boldsymbol{\beta}} = \arg\min_{\boldsymbol{\beta}}
    \left\Vert
    \hat{\boldsymbol{\nu}} - \boldsymbol{\nu}(\hat{\boldsymbol{\alpha}},\boldsymbol{\beta})
    \right\Vert_{\mathbf{W}_{\boldsymbol{\nu}}}^2,
\end{equation}
where $\mathbf{W}_{\boldsymbol{\nu}}$ denotes the Doppler-domain weighting matrix. Using the same first-order linearization as above, the delay-based WLS estimator in~\eqref{sec3-eq13} yields
\begin{equation}\label{sec3-eq15}
    \Delta\hat{\boldsymbol{\alpha}} = 
    \underbrace{
    \left(
    \mathbf{G}_{\boldsymbol{\tau}\boldsymbol{\alpha}}^T\mathbf{W}_{\boldsymbol{\tau}}\mathbf{G}_{\boldsymbol{\tau}\boldsymbol{\alpha}}
    \right)^{-1}
    \mathbf{G}_{\boldsymbol{\tau}\boldsymbol{\alpha}}^T\mathbf{W}_{\boldsymbol{\tau}}}_{\mathbf{E}_{\boldsymbol{\tau}}}
    \Delta\boldsymbol{\tau}.
\end{equation}
Linearizing the Doppler-based WLS estimator in~\eqref{sec3-eq14} yields
\begin{equation}\label{sec3-eq16}
    \Delta\hat{\boldsymbol{\beta}} = \underbrace{
    \left(
    \mathbf{G}_{\boldsymbol{\nu}\boldsymbol{\beta}}^T\mathbf{W}_{\boldsymbol{\nu}}\mathbf{G}_{\boldsymbol{\nu}\boldsymbol{\beta}}
    \right)^{-1}
    \mathbf{G}_{\boldsymbol{\nu}\boldsymbol{\beta}}^T\mathbf{W}_{\boldsymbol{\nu}}}_{\mathbf{E}_{\boldsymbol{\nu}}}
    \left(
    \Delta\boldsymbol{\nu} - \mathbf{G}_{\boldsymbol{\nu}\boldsymbol{\alpha}}\Delta\hat{\boldsymbol{\alpha}}
    \right).
\end{equation}
Let $\hat{\boldsymbol{\xi}}_{\mathrm{SDE}}$ denote the $P$-snapshot sequentially decoupled two-step estimator, and $\Delta\hat{\boldsymbol{\xi}}_{\mathrm{SDE}} \triangleq \big[\Delta\hat{\boldsymbol{\alpha}}^T,\Delta\hat{\boldsymbol{\beta}}^T\big]^T$ denote the estimation error. It follows from~\eqref{sec3-eq15} and~\eqref{sec3-eq16} that the first-order error propagation of $\hat{\boldsymbol{\xi}}_{\mathrm{SDE}}$ can be expressed as $\Delta\hat{\boldsymbol{\xi}}_{\mathrm{SDE}} = \mathbf{E}_{\mathrm{dec}}\Delta\boldsymbol{\zeta}$, where
\begin{equation}\label{sec3-eq17}
    \mathbf{E}_{\mathrm{dec}}
    =
    \begin{bmatrix}
        \mathbf{E}_{\boldsymbol{\tau}} & \mathbf{0} \\
        -\mathbf{E}_{\boldsymbol{\nu}}\mathbf{G}_{\boldsymbol{\nu}\boldsymbol{\alpha}}\mathbf{E}_{\boldsymbol{\tau}} & \mathbf{E}_{\boldsymbol{\nu}}
    \end{bmatrix},
\end{equation}
which leads to 
\begin{equation}\label{sec3-eq18}
    \lim_{P\rightarrow\infty}\mathbf{\Sigma}(\hat{\boldsymbol{\xi}}_{\mathrm{SDE}}) = \mathbf{E}_{\mathrm{dec}}\mathbf{F}^{-1}(\boldsymbol{\zeta})\mathbf{E}_{\mathrm{dec}}^T.
\end{equation}
Compared with the Fisher-optimal local error propagation characterized by $\mathbf{E}_{\mathrm{opt}}$, the sequentially decoupled receiver architecture imposes the structured propagation matrix $\mathbf{E}_{\mathrm{dec}}$. In particular, the zero Doppler-to-position block in its first block renders the position-related estimates locally insensitive to the Doppler residuals, thereby excluding the position-related information available in the Doppler estimates from the position update. The following proposition provides an analytical comparison of the asymptotic covariance matrices to further quantify the impact of this structural restriction.
\begin{proposition}\label{proposition_2}
Under the asymptotic efficiency of the first-stage synchronization-parameter estimator and the full-rank conditions assumed in the above analysis, the asymptotic covariance matrices of the sequentially decoupled and fully coupled two-step estimators satisfy
\begin{equation}\label{sec3-eq19}
    \lim_{P\rightarrow\infty}\left(
    \mathbf{\Sigma}(\hat{\boldsymbol{\xi}}_{\mathrm{SDE}}) - 
    \mathbf{\Sigma}(\hat{\boldsymbol{\xi}}_{\mathrm{FCE}})
    \right) = 
    \Delta\mathbf{E}\;\mathbf{F}^{-1}(\boldsymbol{\zeta})\Delta\mathbf{E}^T \succeq \mathbf{0},
\end{equation}
where $\Delta\mathbf{E} = \mathbf{E}_{\mathrm{dec}} - \mathbf{E}_{\mathrm{opt}}$. Equality is achieved if and only if $\Delta\mathbf{E} = \mathbf{0}$, or equivalently, $\mathbf{E}_{\mathrm{dec}} = \mathbf{E}_{\mathrm{opt}}$. Due to the asymptotic equivalence between the fully coupled estimator and DPE as shown in~\eqref{sec3-eq12}, the covariance gap in~\eqref{sec3-eq19} also applies to DPE.
\end{proposition}
\begin{IEEEproof}
    Please refer to Appendix~\ref{appendix_2}.
\end{IEEEproof}
\begin{remark}\label{remark_2}
    \textit{(Impact of Receiver Architecture)}
    It follows from the above analysis that the Fisher information available from the received observation for estimating the synchronization parameters is entirely characterized by $\mathbf{F}(\boldsymbol{\zeta})$ and is independent of the subsequent receiver architecture. In contrast, the receiver architecture determines how the corresponding synchronization-parameter estimation errors propagate to the receiver state through its local error-propagation matrix. The results in \textbf{Proposition~\ref{proposition_2}} further indicate that the performance loss of the sequentially decoupled receiver arises from its constrained utilization of the available estimation information rather than from the two-step decomposition itself.
\end{remark}
\section{Cram\'er-Rao Bound Analysis}\label{section_4}

\subsection{CRB with Perfect Channel Knowledge}
Under the assumption that the receiver has perfect channel knowledge for the reception of GNSS signals, the bound is obtained by first conditioning on each channel realization and then taking the expectation with respect to the channel statistics~\cite{sec2-ref4}. In this context, the received observation associated with the $k$-th SV obeys a complex Gaussian distribution of  $\mathbf{y}_k\,|\,\boldsymbol{\zeta}_k,\tilde{\mathbf{h}}_k\sim\mathcal{CN}\big(\mathbf{S}_k\tilde{\mathbf{h}}_k,\sigma_{z,k}^2\mathbf{I}_{PN}\big)$. Accordingly, the log-likelihood function for estimating $\boldsymbol{\zeta}_k$ from $\mathbf{y}_k$ is given by
\begin{equation}
    \begin{aligned}\label{sec4-eq1}
        &\ln p(\mathbf{y}_k\,|\,\boldsymbol{\zeta}_k,\tilde{\mathbf{h}}_k) = -P N\ln(\pi\sigma_{z,k}^2)-\frac{1}{\sigma_{z,k}^2}\left\Vert\mathbf{y}_k - \mathbf{S}_k\tilde{\mathbf{h}}_k\right\Vert^2.
    \end{aligned}
\end{equation}
By representing the signal component as $\left(\mathbf{H}_k\otimes\mathbf{I}_N\right)\mathbf{s}_k$, where $\mathbf{H}_k \triangleq \mathrm{diag}(\tilde{\mathbf{h}}_k)$, the Jacobian matrix of the conditional mean with respect to $\boldsymbol{\zeta}_k$ can be written as $\left(\mathbf{H}_k\otimes\mathbf{I}_N\right)\mathbf{J}_k$, where $\mathbf{J}_k\triangleq\frac{\partial\mathbf{s}_k}{\partial\boldsymbol{\zeta}_k^T} = \begin{bmatrix}
    \frac{\partial\mathbf{s}_k}{\partial\tau_k} & \frac{\partial\mathbf{s}_k}{\partial\nu_k}
\end{bmatrix}\in\mathbb{C}^{PN\times2}$. The per-SV FIM conditioned on $\tilde{\mathbf{h}}_k$ is then obtained as
\begin{equation}\label{sec4-eq2}
    \mathbf{F}_k(\boldsymbol{\zeta}_k\,|\,\tilde{\mathbf{h}}_k) = \frac{2}{\sigma_{z,k}^2}\Re\left\{
    \mathbf{J}_k^H\left(
    \mathbf{H}_k^H\mathbf{H}_k
    \otimes\mathbf{I}_N
    \right)\mathbf{J}_k
    \right\},
\end{equation}
Then, averaging the conditional FIM over all possible channel realizations yields
\begin{equation}\label{sec4-eq3}
    \begin{aligned}
        &\mathbf{F}_k^{\mathrm{perf}}(\boldsymbol{\zeta}_k)\triangleq\mathbb{E}_{\tilde{\mathbf{h}}_k}
        \bigg[
        \mathbf{F}_k(\boldsymbol{\zeta}_k\,|\,\tilde{\mathbf{h}}_k)
        \bigg]\\
        &=\frac{2}{\sigma_{z,k}^2}\Re\left\{
        \mathbf{J}_k^H\left(
        \mathbb{E}\left[\mathbf{H}_k^H\mathbf{H}_k\right]
        \otimes\mathbf{I}_N
        \right)\mathbf{J}_k
        \right\} 
        \overset{(a)}{=}\frac{2P_{\tilde{h},k}}{\sigma_{z,k}^2}\Re\left\{\mathbf{J}_k^H\mathbf{J}_k\right\},
    \end{aligned}
\end{equation}
where $(a)$ can be obtained in light of the equality that $\mathbb{E}\left[\mathbf{H}_k^H\mathbf{H}_k\right] = \mathbb{E}\big[\mathrm{diag}(\vert\tilde{\mathbf{h}}_k\vert^2)\big] = P_{\tilde{h},k}\mathbf{I}_{P}$, where $P_{\tilde{h},k}$ denotes the average power gain associated with the $k$-th SV, given as $P_{\tilde{h},k}\triangleq\mathbb{E}[\vert \tilde{h}_{p,k}\vert^2]=\vert\mu_{\tilde{h},k}\vert^2+\sigma_{\tilde{h},k}^2$. Equation~\eqref{sec4-eq3} separates the channel-dependent scaling from the waveform-dependent information, while the latter remains implicit in $\Re\left\{\mathbf{J}_k^H\mathbf{J}_k\right\}$. To gain more useful insights, a closed-form characterization of this Gram matrix is provided in the following proposition.

\begin{proposition}\label{proposition_3}
    Define $\gamma_k\triangleq \frac{P_{\tilde{h},k}}{\sigma_{z,k}^2}$ for notational simplicity. Considering a uniform temporal-energy distribution over the observation interval and a uniform power spectral density for the signal of interest, the channel-averaged conditional FIM associated with the $k$-th SV admits a closed-form expression
    \begin{equation}\label{sec4-eq4}
        \mathbf{F}_k^{\mathrm{perf}}(\boldsymbol{\zeta}_k) =
        8\pi^2PE_s\gamma_k\mathbf{M},
    \end{equation}
    where 
    \begin{equation}\label{sec4-eq5}
        \mathbf{M}=
        \begin{bmatrix}
            B_{\mathrm{eq}}^2/12 +  f_c^2& -f_cT_{\mathrm{obs}}/2\\
            -f_cT_{\mathrm{obs}}/2 & T_{\mathrm{obs}}^2/3
        \end{bmatrix},
    \end{equation}
    with $T_{\mathrm{obs}} = PNT_{\mathrm{samp}}$ denoting the total observation duration, and $B_{\mathrm{eq}}$ denoting the equivalent bandwidth.
\end{proposition}

\begin{IEEEproof}
    Please refer to Appendix~\ref{appendix_3}.
\end{IEEEproof}
Following the PRN-based signal separation, the received observations from different SVs are independent, such that their information contributions are additive. Therefore, substituting~\eqref{sec4-eq4} into~\eqref{sec3-eq2} yields the receiver-state FIM. The corresponding CRB can be then obtained as
\begin{equation}\label{sec4-eq6}
    \mathrm{CRB}_{\boldsymbol{\xi}}^{\mathrm{perf}} = \frac{1}{8\pi^2PE_s}\left[\sum_{k\in\mathcal{K}}\gamma_k\mathbf{G}_k^T\mathbf{M}\mathbf{G}_k\right]^{-1}.\end{equation}
Note that the bound in~\eqref{sec4-eq6} corresponds to the modified CRB defined in~\cite{sec4-ref3}, obtained by inverting the channel-averaged conditional FIM. It should be distinguished from the channel-averaged conditional CRB, which averages the inverse conditional FIM over all possible channel realizations, i.e.,
\begin{equation}\label{sec4-eq7}
    \mathbb{E}_{\tilde{\mathbf{h}}_{\mathcal{K}}}\left[
    \left(
    \sum_{k\in\mathcal{K}} \mathbf{F}_k(\boldsymbol{\xi}\,|\,\tilde{\mathbf{h}}_k)
    \right)^{-1}
    \right]
    \succeq\underbrace{
    \left(
    \sum_{k\in\mathcal{K}}
    \mathbb{E}_{\tilde{\mathbf{h}}_k}
    \bigg[
     \mathbf{F}_k(\boldsymbol{\xi}\,|\,\tilde{\mathbf{h}}_k)
    \bigg]
    \right)^{\!\!-1}}_{\text{modified CRB}},
\end{equation}
where $\tilde{\mathbf{h}}_{\mathcal{K}}\triangleq\big\{\tilde{\mathbf{h}}_k\big\}_{k\in\mathcal{K}}$ denotes the collection of the channel realizations associated with all participating SVs. Due to the inequality in~\eqref{sec4-eq7}, the modified CRB provides an optimistic bound relative to the channel-averaged conditional CRB.

\subsection{CRB with Statistical Channel Knowledge}
When only the statistical channel information is available at the receiver, the instantaneous channel coefficients are treated as latent random variables and marginalized out of the conditional likelihood~\cite{sec4-ref2}. Therefore, the marginal probability density function (PDF) is given by
\begin{equation}
    \begin{aligned}\label{sec4-eq8}
        p(\mathbf{y}_k\,|\,\boldsymbol{\zeta}_k) &= \int
        p\big(
        \mathbf{y}_k\,|\,\boldsymbol{\zeta}_k,\tilde{\mathbf{h}}_k \big)
        p\big(\tilde{\mathbf{h}}_k\big)\,d\tilde{\mathbf{h}}_k\\
        &=\frac{\exp\left(
        -(\mathbf{y}_k - \mathbf{u}_k)^H\mathbf{\Sigma}_k^{-1}(\mathbf{y}_k - \mathbf{u}_k)
        \right)}{\pi^{PN}\det\mathbf{\Sigma}_k},
    \end{aligned}
\end{equation}
where $\mathbf{u}_k = \mu_{\tilde{h},k}\mathbf{s}_k$ denotes the mean of the vector $\mathbf{y}_k$ after marginalization and $\mathbf{\Sigma}_k = \sigma_{z,k}^2\mathbf{I}_{PN} + \sigma_{\tilde{h},k}^2\mathbf{S}_k\mathbf{S}_k^H$ denotes the corresponding covariance matrix. Since both the mean and the covariance depend on $\boldsymbol{\zeta}_k$ through the received waveform, the marginal FIM contains information contributed by variations in both quantities. Using the Slepian-Bang's formula~\cite{sec4-ref4}, the entry at the $u$-th row and $v$-th column of the per-SV marginal FIM under statistical channel knowledge is given by
\begin{equation}
    \begin{aligned}\label{sec4-eq9}
        \left[\mathbf{F}_k^{\mathrm{stat}}(\boldsymbol{\zeta}_k)\right]_{u,v} &= 
        2\Re\left\{
        \frac{\partial\mathbf{u}_k^H}{\partial\zeta_{k,u}}\mathbf{\Sigma}_k^{-1} \frac{\partial\mathbf{u}_k}{\partial\zeta_{k,v}}
        \right\} \\
        &\quad\quad+ \mathrm{tr}\left(
        \mathbf{\Sigma}_k^{-1}\frac{\partial\mathbf{\Sigma}_k}{\partial\zeta_{k,u}}
        \mathbf{\Sigma}_k^{-1}\frac{\partial\mathbf{\Sigma}_k}{\partial\zeta_{k,v}}
        \right).
    \end{aligned}
\end{equation}
Since $\mathbf{S}_k^H\mathbf{S}_k = E_s\mathbf{I}_P$, define $\mathbf{P}_k\triangleq\frac{1}{E_s}\mathbf{S}_k\mathbf{S}_k^H$ as the orthogonal projection matrix onto $\mathcal{S}_k$, and define $\mathbf{P}^\perp_k\triangleq\mathbf{I}_{PN} - \mathbf{P}_k$ as the corresponding projector onto $\mathcal{S}_k^\perp$. Therefore, the covariance matrix and its inverse can be expressed as
\begin{equation}\label{sec4-eq10}
    \mathbf{\Sigma}_k = \sigma_{z,k}^2\mathbf{P}_k^\perp + \big(
    \sigma_{z,k}^2 + E_s\sigma_{\tilde{h},k}^2
    \big )\mathbf{P}_k,
\end{equation}
and
 \begin{equation}\label{sec4-eq11}
    \mathbf{\Sigma}_k^{-1} = \frac{1}{\sigma_{z,k}^2}\mathbf{P}_k^\perp + 
    \frac{1}{ \sigma_{z,k}^2 + E_s\sigma_{\tilde{h},k}^2}
    \mathbf{P}_k,
\end{equation}
respectively. Accordingly, the Fisher information contribution associated with the coherent mean can be expressed as
\begin{equation}
    \begin{aligned}\label{sec4-eq12}
        \left[\mathbf{F}_{k}^{\mathrm{mean}}\right]_{u,v} &=
        \frac{2\vert\mu_{\tilde{h},k}\vert^2}{\sigma_{z,k}^2}\Re\left\{
        \frac{\partial\mathbf{s}_k^H}{\partial\zeta_{k,u}}\mathbf{P}_k^\perp
        \frac{\partial\mathbf{s}_k}{\partial\zeta_{k,v}}
        \right\}\\
        &+\frac{2\vert\mu_{\tilde{h},k}\vert^2}{\sigma_{z,k}^2+E_s\sigma_{\tilde{h},k}^2}\Re\left\{
        \frac{\partial\mathbf{s}_k^H}{\partial\zeta_{k,u}}\mathbf{P}_k
        \frac{\partial\mathbf{s}_k}{\partial\zeta_{k,v}}
        \right\}.
    \end{aligned}
\end{equation}
We next simplify the term within the trace operator in~\eqref{sec4-eq9}. It follows from~\eqref{sec4-eq10} that $\frac{\partial\mathbf{\Sigma}_k}{\partial\zeta_{k,u}} = E_s\sigma_{\tilde{h},k}^2\frac{\partial\mathbf{P}_k}{\partial\zeta_{k,u}}$. Substituting this result into the inverse covariance in~\eqref{sec4-eq11} yields
\begin{equation}\label{sec4-eq13}
    \mathbf{\Sigma}_k^{-1}\frac{\partial\mathbf{\Sigma}_k}{\partial\zeta_{k,u}} = 
    E_s\sigma_{\tilde{h},k}^2\left(
    \frac{\mathbf{P}_k^\perp}{\sigma_{z,k}^2} + 
    \frac{\mathbf{P}_k}{ \sigma_{z,k}^2 + E_s\sigma_{\tilde{h},k}^2}
    \right)\frac{\partial\mathbf{P}_k}{\partial\zeta_{k,u}}.
\end{equation}
After some straightforward manipulations, we have
\begin{align}\label{sec4-eq14}
    &\mathbf{P}_k^\perp\frac{\partial\mathbf{P}_k}{\partial\zeta_{k,u}} = \frac{1}{E_s}\mathbf{P}_k^\perp\frac{\partial\mathbf{S}_{k}}{\partial\zeta_{k,u}}\mathbf{S}_k^H,&
    \mathbf{P}_k\frac{\partial\mathbf{P}_k}{\partial\zeta_{k,u}} &= \frac{1}{E_s}\mathbf{S}_k\frac{\partial\mathbf{S}_k^H}{\partial\zeta_{k,u}}\mathbf{P}_k^\perp.
\end{align}
Substituting~\eqref{sec4-eq13} and~\eqref{sec4-eq14} into the covariance-dependent term in~\eqref{sec4-eq9} yields              
\begin{equation}
    \begin{aligned}\label{sec4-eq15}
        \left[\mathbf{F}_{k}^{\mathrm{cov}}\right]_{u,v} &=\frac{2E_s\sigma_{\tilde{h},k}^4}{\sigma_{z,k}^2\left(\sigma_{z,k}^2+E_s\sigma_{\tilde{h},k}^2\right)}\Re\left\{
        \mathrm{tr}\left(
        \frac{\partial\mathbf{S}_k^H}{\partial\zeta_{k,u}}\mathbf{P}_k^\perp\frac{\partial\mathbf{S}_k}{\partial\zeta_{k,v}}
        \right)
        \right\}\\
        &=\frac{2E_s\sigma_{\tilde{h},k}^4}{\sigma_{z,k}^2\left(\sigma_{z,k}^2+E_s\sigma_{\tilde{h},k}^2\right)}\Re\left\{
        \frac{\partial\mathbf{s}_k^H}{\partial\zeta_{k,u}}\mathbf{P}_k^\perp
        \frac{\partial\mathbf{s}_k}{\partial\zeta_{k,v}}
        \right\},
    \end{aligned}
\end{equation}
where the last equality follows by exploiting the corresponding block-diagonal structures of $\mathbf{S}_k$ and $\mathbf{P}_k^\perp$. Finally, combining the results in~\eqref{sec4-eq12} and~\eqref{sec4-eq15} yields the simplified entry-wise expression of $\mathbf{F}_k^{\mathrm{stat}}(\boldsymbol{\zeta}_k)$, and its closed-form representation is given in the following proposition.
\begin{proposition}\label{proposition_4}
    Define $\rho_k\triangleq\frac{E_s\sigma_{\tilde{h},k}^2}{\sigma_{z,k}^2}$ for notational simplicity. Furthermore, we define $\chi_{k,1}\triangleq\frac{\vert\mu_{\tilde{h},k}\vert^2}{P_{\tilde{h},k}}$ and $\chi_{k,2}\triangleq\frac{ \sigma_{\tilde{h},k}^2}{P_{\tilde{h},k}}$ as the fractions of the average channel power contributed by the coherent mean and channel randomness, respectively, with $\chi_{k,1}+\chi_{k,2}=1$. The marginal FIM associated with the $k$-th SV admits a closed-form representation
    \begin{equation}\label{sec4-eq16}
        \mathbf{F}_k^{\mathrm{stat}}(\boldsymbol{\zeta}_k) = 8\pi^2PE_s\gamma_k\left[
        \left(
        1 - \frac{\chi_{k,2}}{1+\rho_k}
        \right)\mathbf{M}_1 + \frac{\chi_{k,1}}{1+\rho_k}\mathbf{M}_2
        \right],
    \end{equation}
    where
    \begin{equation}\label{sec4-eq17}
        \mathbf{M}_1
        = \begin{bmatrix}
            B_{\mathrm{eq}}^2/12 & 0\\
            0 & T_{\mathrm{code}}^2/12
        \end{bmatrix},
    \end{equation}
    and \begin{equation}\label{sec4-eq18}
        \mathbf{M}_2 
        = \begin{bmatrix}
            f_c^2 & -f_cT_{\mathrm{obs}}/2\\
            -f_cT_{\mathrm{obs}}/2 & T_{\mathrm{obs}}^2/3 - T_{\mathrm{code}}^2/12
        \end{bmatrix}.
    \end{equation}
    It can be readily obtained that $\mathbf{M}_1 + \mathbf{M}_2 = \mathbf{M}$.  
\end{proposition}
\begin{IEEEproof}
    Please refer to Appendix~\ref{appendix_4}.
\end{IEEEproof}
\textbf{Proposition~\ref{proposition_4}} shows that the waveform-domain Fisher information can be decomposed into $\mathbf{M}_1$ and $\mathbf{M}_2$, which characterize the contributions associated with the waveform sensitivities projected onto $\mathcal{S}_k^\perp$ and $\mathcal{S}_k$, respectively. Their contributions to the marginal FIM are jointly determined by $\gamma_k$, $\chi_{k,1}$, $\chi_{k,2}$, and $\rho_k$. To distinguish these coupled effects on receiver-state estimation, the following corollaries characterize the asymptotic behavior of the corresponding CRB under two complementary limiting regimes.

\begin{corollary}\label{corollary_1}
    For fixed $\chi_{k,1}$ and $\chi_{k,2}$, the asymptotic expressions of the marginal CRB as $\gamma_k\rightarrow0$ and $\gamma_k\rightarrow\infty$ can be derived as
    \begin{align}
    \lim_{\gamma_k\rightarrow0} \mathrm{CRB}_{\boldsymbol{\xi}}^{\mathrm{stat}} &= 
    \frac{1}{8\pi^2PE_s}\left[\sum_{k\in\mathcal{K}}\gamma_k\chi_{k,1}\mathbf{G}_k^T\mathbf{M}
    \mathbf{G}_k
    \right]^{-1},\label{sec4-eq19}\\
    \lim_{\gamma_k\rightarrow\infty} \mathrm{CRB}_{\boldsymbol{\xi}}^{\mathrm{stat}} &=
    \frac{1}{8\pi^2PE_s}\left[\sum_{k\in\mathcal{K}}\gamma_k\mathbf{G}_k^T\mathbf{M}_1
    \mathbf{G}_k
    \right]^{-1}.\label{sec4-eq20}
    \end{align}
\end{corollary}
\begin{IEEEproof}
    Given that $\rho_k\triangleq\frac{E_s\sigma_{\tilde{h},k}^2}{\sigma_{z,k}^2} = E_s\chi_{k,2}\gamma_k$, we have $(1+\rho_k)^{-1}\approx1-\rho_k$ as $\gamma_k\rightarrow0$. The weighting factors of $\mathbf{M}_1$ and $\mathbf{M}_2$ both approach $\chi_{k,1}$ to the leading order. Since the marginal FIM is further scaled by $\gamma_k$, the remaining terms are $O(\gamma_k^2)$ and can therefore be neglected. Hence, its leading-order approximation is proportional to $\gamma_k\chi_{k,1}\mathbf{M}$. Mapping the resulting per-SV FIMs to the receiver state through $\mathbf{G}_k$, summing over the participating SVs, and taking the matrix inverse yields the first asymptotic CRB. As $\gamma_k\rightarrow\infty$ with $\chi_{k,2}>0$, we have $(1+\rho_k)^{-1}\approx\rho_k^{-1}$. Therefore, the weighting factor associated with $\mathbf{M}_1$ is $O(1)$, whereas that associated with $\mathbf{M}_2$ is $O(\gamma_k^{-1})$. After accounting for the overall scaling factor $\gamma_k$ in the marginal FIM, the corresponding information contributions scale as $O(\gamma_k)$ and $O(1)$, respectively. Hence, the $\mathbf{M}_2$ contribution is asymptotically negligible relative to that of $\mathbf{M}_1$, and the leading-order marginal FIM is determined solely by $\mathbf{M}_1$. Applying the same geometric mapping, summation, and matrix inversion gives the second asymptotic CRB. This completes the proof.
\end{IEEEproof}

\begin{corollary}\label{corollary_2}
    For fixed $\gamma_k$, the asymptotic expressions of the marginal CRB as $\chi_{k,2}\rightarrow0$ and $\chi_{k,2}\rightarrow1$ can be derived as
    \begin{align}
        \lim_{\chi_{k,2}\rightarrow0} \mathrm{CRB}_{\boldsymbol{\xi}}^{\mathrm{stat}}
        &=\frac{1}{8\pi^2PE_s}
        \left[\sum_{k\in\mathcal{K}}\gamma_k
        \mathbf{G}_k^T\mathbf{M}\mathbf{G}_k
        \right]^{-1},\label{sec4-eq21}\\
        \lim_{\chi_{k,2}\rightarrow1} \mathrm{CRB}_{\boldsymbol{\xi}}^{\mathrm{stat}}
        &=
        \frac{1}{8\pi^2PE_s}
        \left[\sum_{k\in\mathcal{K}}
        \frac{E_s\gamma_k^2}{1+E_s\gamma_k}
        \mathbf{G}_k^T\mathbf{M}_1\mathbf{G}_k
        \right]^{-1}.\label{sec4-eq22}
    \end{align}
\end{corollary}

\begin{IEEEproof}
    As $\chi_{k,2}\rightarrow0$, we have $\chi_{k,1}\rightarrow1$ and $\rho_k\rightarrow0$. Therefore, the weighting factors of $\mathbf{M}_1$ and $\mathbf{M}_2$ both approach unity, and the marginal FIM thus reduces to $8\pi^2PE_s\gamma_k\mathbf{M}$. Mapping the resulting per-SV FIMs to the receiver state through $\mathbf{G}_k$, summing over the participating SVs, and taking the matrix inverse yields the first asymptotic CRB. As $\chi_{k,2}\rightarrow1$, we have $\chi_{k,1}\rightarrow0$ and $\rho_k\rightarrow E_s\gamma_k$. Therefore, the weighting factor of $\mathbf{M}_2$ approaches zero, whereas that of $\mathbf{M}_1$ approaches $\frac{E_s\gamma_k}{1+E_s\gamma_k}$. Applying the same geometric mapping, summation, and matrix inversion gives the second asymptotic CRB. This completes the proof.
\end{IEEEproof}

\begin{remark}\label{remark_3}
    \textit{(Impact of Channel Uncertainty)}
    \textbf{Corollary~\ref{corollary_1}} illustrates the fundamental impact of channel uncertainty on the waveform-domain information availability. As $\gamma_k\rightarrow\infty$, the effect of AWGN becomes negligible and the intrinsic ambiguity introduced by the unknown instantaneous channel becomes explicit. As discussed in \textbf{Remark~\ref{remark_1}}, the signal component associated with channel randomness is confined to $\mathcal{S}_k$. Therefore, the waveform sensitivities in this subspace cannot be fully distinguished from the random channel contribution. In contrast, those lying within $\mathcal{S}_k^\perp$ remain distinguishable. Consequently, the information associated with $\mathbf{M}_1$ continues to increase with $\gamma_k$, whereas that associated with $\mathbf{M}_2$ remains bounded by channel ambiguity. It should be noted that as $\gamma_k\rightarrow0$, the random channel contribution becomes negligible relative to AWGN, such that $\mathbf{\Sigma}_k\approx\sigma_{z,k}^2\mathbf{I}_{PN}$. As a result, the distinction between $\mathcal{S}_k$ and $\mathcal{S}_k^\perp$ vanishes, and the Fisher information retains the complete waveform-information structure.
\end{remark}
\begin{remark}\label{remark_4}
    \textit{(Different Levels of Randomness)}
    \textbf{Corollary~\ref{corollary_2}} illustrates how the level of channel randomness affects the representation of waveform-domain Fisher information in the observation statistics. As $\chi_{k,2}\rightarrow0$, the channel becomes deterministic and the marginal FIM is determined entirely by the coherent-mean contribution, which remains sensitive to waveform perturbations in both $\mathcal{S}_k$ and $\mathcal{S}_k^\perp$, thereby preserving both $\mathbf{M}_1$ and $\mathbf{M}_2$. As $\chi_{k,2}\rightarrow1$, the coherent mean vanishes and the marginal FIM arises solely from the covariance. Since $\mathbf{\Sigma}_k$ depends on $\boldsymbol{\zeta}_k$ through $\mathbf{P}_k$, any covariance variation induced by a perturbation in $\boldsymbol{\zeta}_k$ must result from a change in $\mathcal{S}_k$ itself. Hence, only waveform sensitivities that alter $\mathcal{S}_k$ can contribute to the Fisher information through the covariance, whereas those confined within $\mathcal{S}_k$ leave $\mathbf{P}_k$ unchanged and provide no covariance-dependent Fisher information. Consequently, only $\mathbf{M}_1$ remains in this limit, while $\mathbf{M}_2$ vanishes. 
\end{remark}

\section{Misspecified Cram\'er-Rao Bound Analysis}\label{section_5}
\subsection{Model Misspecification and Pseudo-True Parameters}
To streamline the analysis, we first define the true receiver state as $\bar{\boldsymbol{\xi}} = \big[\bar{\mathbf{p}}^T,\bar{\mathbf{v}}^T,\bar{\delta t}\big]^T$, with the corresponding true synchronization parameters associated with each SV denoted by $\bar{\boldsymbol{\zeta}}_k = \big[\bar{\tau}_k,\bar{\nu}_k\big]^T$. An overline on a waveform-dependent quantity indicates its evaluation at the true receiver state, i.e., $\bar{\mathbf{s}}_k\triangleq\mathbf{s}_k(\bar{\boldsymbol{\xi}})$ and $\bar{\mathbf{S}}_k\triangleq\mathbf{S}_k(\bar{\boldsymbol{\xi}})$. Marginalizing over the random channel yields the true observation distribution in~\eqref{sec4-eq8}, i.e., $\mathbf{y}_k\,|\,\bar{\boldsymbol{\zeta}}_k\sim\mathcal{CN}\big(\bar{\mathbf{u}}_k,\bar{\mathbf{\Sigma}}_k\big)$, where $\bar{\mathbf{u}}_k = \mu_{\tilde{h},k}\bar{\mathbf{s}}_k$ and $\bar{\mathbf{\Sigma}}_k = \sigma_{z,k}^2\mathbf{I}_{PN} +\sigma_{\tilde{h},k}^2\bar{\mathbf{S}}_k\bar{\mathbf{S}}_k^H$ denote the true mean and true covariance, respectively. In contrast, a practical GNSS receiver may adopt a simplified observation model, in which the channel associated with the $k$-th SV is represented by an unknown deterministic complex gain $a_k$ that remains constant over the $P$ code periods, while the random channel variations are ignored and the residual disturbance is modeled as AWGN with covariance $\sigma_{z,k}^2\mathbf{I}_{PN}$. Accordingly, the assumed observation distribution is given by $\mathbf{y}_k\,|\,\boldsymbol{\zeta}_k,a_k\sim\mathcal{CN}(a_k\mathbf{s}_k,\sigma_{z,k}^2\mathbf{I}_{PN})$.

We next define the augmented parameter vector under the misspecified model as $\tilde{\boldsymbol{\xi}} = \big[\boldsymbol{\xi}^T,a_{R,1},a_{I,1},\ldots,a_{R,K},a_{I,K}\big]^T$, where
$a_{R,k}=\Re\{a_k\}$ and $a_{I,k}=\Im\{a_k\}$ denote the real and imaginary parts of $a_k$, respectively. Let $p(\boldsymbol{y}\,|\,\bar{\boldsymbol{\xi}})$ and $\tilde{p}(\boldsymbol{y}\,|\,\tilde{\boldsymbol{\xi}})$ denote the joint PDFs of the received observations under the true and misspecified models, respectively. Then, the pseudo-true parameter vector that minimizes the Kullback-Leibler (KL) divergence between the true and the misspecified PDFs can be characterized as~\cite{sec5-ref1,sec5-ref2,sec5-ref3}
\begin{equation}\label{sec5-eq1}
    \tilde{\boldsymbol{\xi}}^\star = \arg\min_{\tilde{\boldsymbol{\xi}}}D_{KL}
    \left(
    p(\boldsymbol{y}\,|\,\bar{\boldsymbol{\xi}})\,\big\Vert
    \,\tilde{p}(\boldsymbol{y}\,|\,\tilde{\boldsymbol{\xi}})
    \right).
\end{equation}
Using the standard KL divergence between complex Gaussian distributions and removing the terms independent of $\tilde{\boldsymbol{\xi}}$, the pseudo-true parameter vector can be equivalently given by~\cite{sec5-ref4}
\begin{equation}\label{sec5-eq2}
    \tilde{\boldsymbol{\xi}}^\star = \arg\min_{\tilde{\boldsymbol{\xi}}}
    \sum_{k\in\mathcal{K}}
    \frac{1}{\sigma_{z,k}^2}
    \left\Vert
    \bar{\mathbf{u}}_k - a_k\mathbf{s}_k(\boldsymbol{\xi})
    \right\Vert^2.
\end{equation}
To solve~\eqref{sec5-eq2}, we first minimize the objective with respect to the nuisance complex gains for a fixed candidate receiver state. Using $\Vert\mathbf{s}_k\Vert^2 = PE_s$, the resulting conditional minimizer can be equivalently obtained as
\begin{equation}\label{sec5-eq3}
    a_k^\star(\boldsymbol{\xi
    }) = \mu_{\tilde{h},k}\frac{\mathbf{s}_k^H(\boldsymbol{\xi})\bar{\mathbf{s}}_k}{PE_s}.
\end{equation}
Substituting~\eqref{sec5-eq3} into~\eqref{sec5-eq2} yields 
\begin{equation}
    \boldsymbol{\xi}^\star = \arg\min_{\boldsymbol{\xi}}\sum_{k\in\mathcal{K}}
    \frac{1}{\sigma_{\tilde{h},k}^2}
    \left\Vert
    \mu_{\tilde{h},k}\bar{\mathbf{s}_k} - \frac{\mu_{\tilde{h},k}\mathbf{s}_k^H(\boldsymbol{\xi})\bar{\mathbf{s}}_k\mathbf{s}_k(\boldsymbol{\xi})}{PE_s}
    \right\Vert^2.
\end{equation}
After some straightforward manipulations, we have
\begin{equation}\label{sec5-eq4}
    \boldsymbol{\xi}^\star
    =
    \arg\max_{\boldsymbol{\xi}}\sum_{k\in\mathcal{K}}
        \frac{\vert\mu_{\tilde{h},k}\vert^2}{\sigma_{\tilde{h},k}^2PE_s}
        \Vert
        \mathbf{s}_k^H(\boldsymbol{\xi})\bar{\mathbf{s}_k}
        \Vert^2.
\end{equation}
By using the Cauchy-Schwarz inequality, the objective in~\eqref{sec5-eq4} attains its maximum at the true receiver state. Provided that the waveform characteristics and satellite geometry ensure local identifiability of the receiver state, this maximizer is locally unique, which yields $\boldsymbol{\xi}^\star=\bar{\boldsymbol{\xi}}$ and $a_k^\star=\mu_{\tilde h,k}$. Thus, the pseudo-true parameter vector can be obtained as $\tilde{\boldsymbol{\xi}}^\star = \big[
\bar{\boldsymbol{\xi}}^T,\Re\{\mu_{\tilde{h},1}\},\Im\{\mu_{\tilde{h},1}\},\ldots,\Re\{\mu_{\tilde{h},K}\},\Im\{\mu_{\tilde{h},K}\}
\big]^T$.

\subsection{Derivation of MCRB for the Misspecified Model}
For the $k$-th SV, define the augmented synchronization parameter vector as $\tilde{\boldsymbol{\zeta}}_k
\triangleq
\big[\boldsymbol{\zeta}_k^T,a_{R,k},a_{I,k}\big]^T$. According to the pseudo-true parameters derived above, its pseudo-true value is given by $\tilde{\boldsymbol{\zeta}}_k^\star
=\big[
\bar{\boldsymbol{\zeta}}_k^T,
\Re\{\mu_{\tilde{h},k}\},
\Im\{\mu_{\tilde{h},k}\}
\big]^T$. Then, define the negative expected Hessian and the covariance matrix of the misspecified score as
\begin{align}
    \mathbf{C}_k
    &\triangleq
    -\mathbb{E}_p
    \left[
    \left.
    \nabla_{\tilde{\boldsymbol{\zeta}}_k}^{2}
    \ln\tilde{p}
    (\mathbf{y}_k\,|\,\tilde{\boldsymbol{\zeta}}_k)
    \right|_{\tilde{\boldsymbol{\zeta}}_k
    =\tilde{\boldsymbol{\zeta}}_k^\star}
    \right],
    \label{sec5-eq5}\\
    \mathbf{D}_k
    &\triangleq
    \mathbb{E}_p
    \left[
    \left.
    \nabla_{\tilde{\boldsymbol{\zeta}}_k}
    \ln\tilde{p}
    (\mathbf{y}_k\,|\,\tilde{\boldsymbol{\zeta}}_k)
    \nabla_{\tilde{\boldsymbol{\zeta}}_k}^{T}
    \ln\tilde{p}
    (\mathbf{y}_k\,|\,\tilde{\boldsymbol{\zeta}}_k)
    \right|_{\tilde{\boldsymbol{\zeta}}_k
    =\tilde{\boldsymbol{\zeta}}_k^\star}
    \right].
    \label{sec5-eq6}
\end{align}
It should be noted that all derivatives in the above equations are taken with respect to the candidate parameters before evaluation at $\tilde{\boldsymbol{\zeta}}_k^\star$, whereas the expectation is taken under the true distribution. In the following derivations, all waveform quantities and geometric Jacobians are evaluated at $\bar{\boldsymbol{\zeta}}_k$, and the overlines are omitted for notational simplicity. Under independent reception from different SVs, the corresponding log-likelihood function of the received observation associated with the misspecified model is given by
\begin{equation}\label{sec5-eq7}
    \ln\tilde{p}
    (\mathbf{y}_k\,|\,\tilde{\boldsymbol{\zeta}}_k)
    =
    -PN\ln(\pi\sigma_{z,k}^{2})
    -
    \frac{1}{\sigma_{z,k}^{2}}
    \left\|
    \mathbf{y}_k-a_k\mathbf{s}_k
    \right\|^2.
\end{equation}
Substituting~\eqref{sec5-eq7} into~\eqref{sec5-eq5} and~\eqref{sec5-eq6}, the true and assumed mean vectors coincide. Consequently, the terms
involving the second-order derivatives of the assumed mean
vanish after taking the expectation, which leads to
\begin{align}\label{sec5-eq8}
    &\mathbf{C}_k = \frac{2}{\sigma_{z,k}^2}\Re\left\{
    \tilde{\mathbf{J}}_k^H\tilde{\mathbf{J}}_k
    \right\}, & 
    \mathbf{D}_k = \frac{2}{\sigma_{z,k}^4}
    \Re\left\{
    \tilde{\mathbf{J}}_k^H\mathbf{\Sigma}_k\tilde{\mathbf{J}}_k
    \right\},
\end{align}
where $\tilde{\mathbf{J}}_k\triangleq\left.\frac{\partial(a_k\mathbf{s}_k)}{\partial\tilde{\boldsymbol{\zeta}}_k^T}\right|_{\tilde{\boldsymbol{\zeta}}_k = \tilde{\boldsymbol{\zeta}}_k^\star} = \begin{bmatrix}
    \mu_{\tilde{h},k}\mathbf{J}_k&\mathbf{s}_k&\jmath\mathbf{s}_k
\end{bmatrix}$ denotes the Jacobian of the assumed mean with respect to $\tilde{\boldsymbol{\zeta}}_k$.

To characterize the estimation performance associated with $\boldsymbol{\zeta}_k$, we treat $\mathbf{a}_k\triangleq\big[a_{R,k},a_{I,k}\big]^T$ as nuisance parameters. Accordingly, $\mathbf{C}_k$ can be partitioned as
\begin{equation}\label{sec5-eq9}
    \mathbf{C}_k = \begin{bmatrix}
        \mathbf{C}_{k,\boldsymbol{\zeta}\boldsymbol{\zeta}} & 
        \mathbf{C}_{k,\boldsymbol{\zeta}\mathbf{a}}\\[0.5ex] 
        \mathbf{C}_{k,\boldsymbol{\zeta}\mathbf{a}}^T &
        \mathbf{C}_{k,\mathbf{a}\mathbf{a}}
    \end{bmatrix},
\end{equation}
where
\begin{align}
    &\mathbf{C}_{k,\boldsymbol{\zeta}\boldsymbol{\zeta}} = \frac{2\vert
    \mu_{\tilde{h},k}
    \vert^2}{\sigma_{z,k}^2}\Re\left\{
    \mathbf{J}_{k}^H\mathbf{J}_{k}
    \right\},\label{sec5-eq10}\\
    &\mathbf{C}_{k,\mathbf{a}\mathbf{a}} = \frac{2PE_s}{\sigma_{z,k}^2}\mathbf{I}_{2},\label{sec5-eq11}\\
    &\mathbf{C}_{k,\boldsymbol{\zeta}\mathbf{a}} = \frac{2}{\sigma_{z,k}^2}
    \Re\left\{
    \mu_{\tilde{h},k}^*\mathbf{J}_k^H\begin{bmatrix}
        \mathbf{s}_k & \jmath\mathbf{s}_k
    \end{bmatrix}
    \right\}.\label{sec5-eq12}
\end{align}
Using the Schur complement, the equivalent negative expected Hessian for the synchronization parameters is given by
\begin{equation}\label{sec5-eq13}
    \mathbf{C}_k^e = \mathbf{C}_{k,\boldsymbol{\zeta}\boldsymbol{\zeta}} -
    \mathbf{C}_{k,\boldsymbol{\zeta}\mathbf{a}}\mathbf{C}_{k,\mathbf{a}\mathbf{a}}^{-1}\mathbf{C}_{k,\boldsymbol{\zeta}\mathbf{a}}^T.
\end{equation}
Let $\mathbf{L}_k\triangleq\begin{bmatrix}
        \mathbf{I}_2 & -\mathbf{C}_{k,\boldsymbol{\zeta}\mathbf{a}}\mathbf{C}_{k,\mathbf{a}\mathbf{a}}^{-1}
    \end{bmatrix}$, the corresponding equivalent score covariance is then given by
\begin{equation}\label{sec5-eq14}
    \mathbf{D}_k^e = \mathbf{L}_k\mathbf{D}_k\mathbf{L}_k^T.
\end{equation}
To facilitate the subsequent derivation of the receiver-state MCRB, we first establish the closed-form expressions for $\mathbf{C}_k^e$ and $\mathbf{D}_k^e$ in the following proposition.

\begin{proposition}\label{proposition_5}
    Under the same uniform temporal-energy and uniform power spectrum approximations, the closed-form expressions for
    $\mathbf{C}_k^e$ and $\mathbf{D}_k^e$ are given by
    \begin{align}
        \mathbf{C}_k^e
        &=
        8\pi^2PE_s\gamma_k\chi_{k,1}\mathbf{M}_3,
        \label{sec5-eq15}\\
        \mathbf{D}_k^e
        &=
        8\pi^2PE_s\gamma_k\chi_{k,1}
        \left[
        \mathbf{M}_3 + \rho_k(\mathbf{M}_3-\mathbf{M}_1)
        \right],
        \label{sec5-eq16}
    \end{align}
    where
    \begin{equation}
        \mathbf{M}_3
        \triangleq
        \begin{bmatrix}
            B_{\rm eq}^2/12 & 0\\
            0 & T_{\rm obs}^2/12
        \end{bmatrix}.\label{sec5-eq17}
    \end{equation}
\end{proposition}
\begin{IEEEproof}
    Please refer to Appendix~\ref{appendix_5}.
\end{IEEEproof}

Using the relation between the synchronization parameters and the receiver state, the equivalent synchronization-parameter score is mapped to the receiver state through the Jacobian $\mathbf{G}_k$. Since the expected score vanishes at the pseudo-true parameters, the corresponding expected Hessian follows the associated transformation, which yields
\begin{equation}\label{sec5-eq18}
    \mathbf{C}_{\boldsymbol{\xi}}^e
    =
    8\pi^2PE_s
    \sum_{k\in\mathcal{K}}
    \gamma_k\chi_{k,1}
    \mathbf{G}_k^T
    \mathbf{M}_3
    \mathbf{G}_k,
\end{equation}
and
\begin{equation}\label{sec5-eq19}
    \mathbf{D}_{\boldsymbol{\xi}}^e
    =
    8\pi^2PE_s
    \sum_{k\in\mathcal{K}}
    \gamma_k\chi_{k,1}
    \mathbf{G}_k^T
    \left[
    \mathbf{M}_3 + \rho_k(\mathbf{M}_3-\mathbf{M}_1)
    \right]
    \mathbf{G}_k.
\end{equation}
Equivalently, $\mathbf{D}_{\boldsymbol{\xi}}^e$ can be decomposed as
\begin{equation}\label{sec5-eq20}
    \mathbf{D}_{\boldsymbol{\xi}}^e
    =
    \mathbf{C}_{\boldsymbol{\xi}}^e
    +\
    \underbrace{
    8\pi^2PE_s
    \sum_{k\in\mathcal{K}}
    \gamma_k\chi_{k,1}\rho_k
    \mathbf{G}_k^T
    \left(\mathbf{M}_3-\mathbf{M}_1\right)
    \mathbf{G}_k}_{\triangleq\, \Delta\mathbf{D}^e_{\boldsymbol{\xi}}},
\end{equation}
where $ \Delta\mathbf{D}^e_{\boldsymbol{\xi}}$
characterizes the additional score covariance induced by the unmodeled random channel variations. Consequently, the receiver-state MCRB is given by
\begin{equation}
    \begin{aligned}
        \mathrm{CRB}_{\boldsymbol{\xi}}^{\mathrm{miss}}
        &=
        \left(\mathbf{C}_{\boldsymbol{\xi}}^e\right)^{-1}
        \mathbf{D}_{\boldsymbol{\xi}}^e
        \left(\mathbf{C}_{\boldsymbol{\xi}}^e\right)^{-1}\\
        &=
        \left(\mathbf{C}_{\boldsymbol{\xi}}^e\right)^{-1}
        +
        \left(\mathbf{C}_{\boldsymbol{\xi}}^e\right)^{-1}
        \Delta\mathbf{D}^e_{\boldsymbol{\xi}}
        \left(\mathbf{C}_{\boldsymbol{\xi}}^e\right)^{-1}.\label{sec5-eq22}
    \end{aligned}
\end{equation}

\begin{remark}\label{remark_5}
    \textit{(Impact of Model Misspecification)}
    Under the considered channel model where the per-SV complex gain is treated as an unknown nuisance parameter while the random channel variations across adjacent code periods are neglected, the effect of model misspecification is entirely characterized by $ \Delta\mathbf{D}^e_{\boldsymbol{\xi}}$. For a fixed $\gamma_k$, when the channel randomness vanishes, i.e., $\chi_{k,2}\rightarrow0$, we have $\rho_k\rightarrow0$, which yields $\Delta\mathbf{D}^e_{\boldsymbol{\xi}}\rightarrow\mathbf{0}$. In this case, the assumed model becomes correctly specified, and the MCRB reduces to the conventional CRB, i.e.,
    \begin{equation}\label{sec5-eq23}
        \lim_{\chi_{k,2}\rightarrow0}\mathrm{CRB}_{\boldsymbol{\xi}}^{\mathrm{miss}} = \frac{1}{8\pi^2PE_s}\left[
        \sum_{k\in\mathcal{K}}\gamma_k\mathbf{G}_k^T\mathbf{M}_3\mathbf{G}_k
        \right]^{-1}.
    \end{equation}
    Compared with~\eqref{sec4-eq21}, where the channel randomness also vanishes but the resulting deterministic channel gain is known, the CRB in~\eqref{sec5-eq23} is characterized by $\mathbf{M}_3$ rather than the complete $\mathbf{M}$. This is because when the complex channel coefficient associated with each SV is completely unknown and therefore must be eliminated as a nuisance parameter, the waveform sensitivities that are indistinguishable from variations of this complex gain cannot contribute to the Fisher information associated with the receiver state, thereby resulting in an information loss characterized by $\mathbf{M}-\mathbf{M}_3$.
\end{remark}

\begin{figure}[hb]
    \centering
    \includegraphics[width=0.85\linewidth]{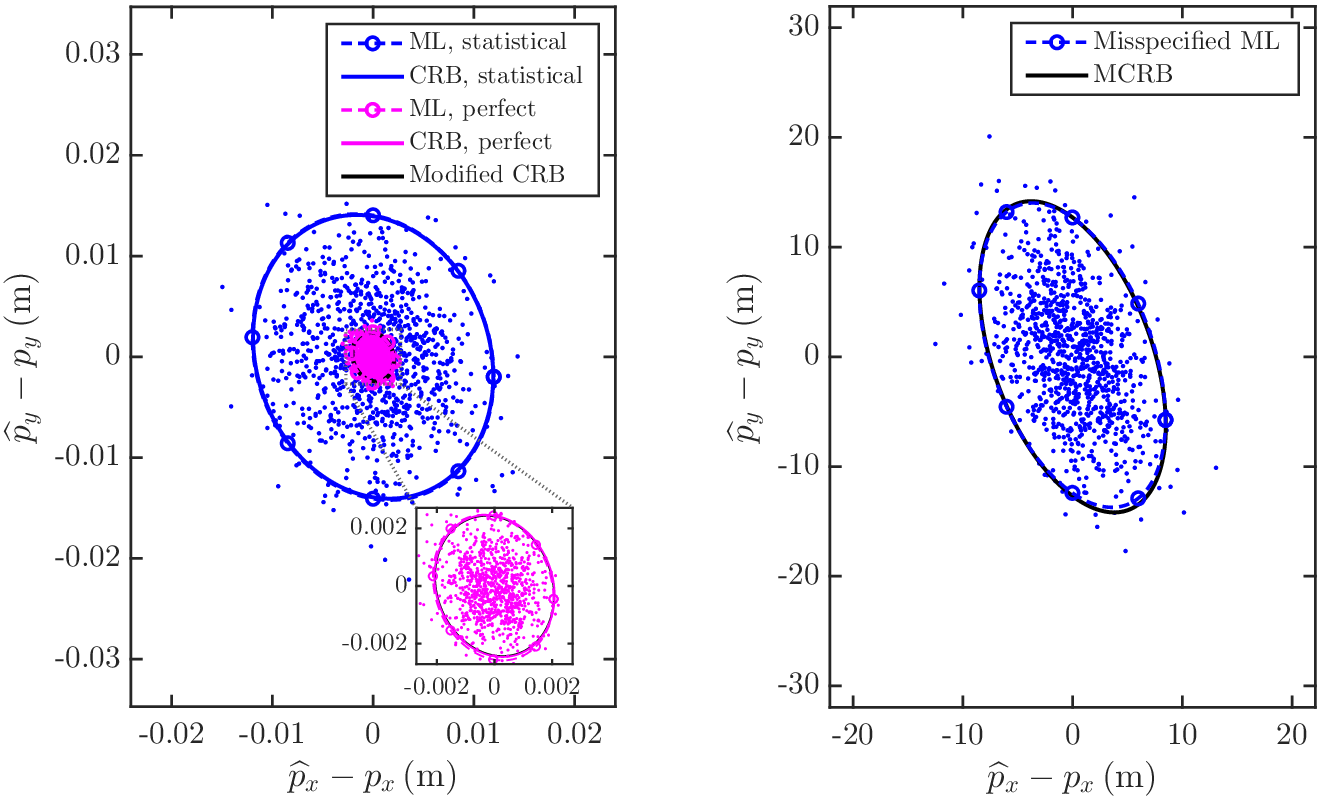}
    \caption{Comparison of $95\%$ horizontal positioning error ellipses obtained from empirical ML estimates and the derived CRBs and MCRB.}
    \label{figure_1}
\end{figure}
\section{Numerical Results}\label{section_6}
This section provides numerical results to validate the derived performance bounds. Unless otherwise specified, we set $f_c = 1575.42$ MHz, $R_{\mathrm{chip}} = 1.023$ MHz, $T_{\mathrm{code}} = 1$ ms, $T_{\mathrm{samp}} = 200$ ns, $T_{\mathrm{obs}} = 50$ ms, $E_s = N = 5000$, and $\chi_{k,1} = \chi_{k,2} = 0.5$. The carrier-to-noise ratio is set to be $C/N_0 = 45$ dB-Hz. The receiver position and velocity are set to be $\bar{\mathbf{p}} = [5,8,10]^T$ and $\bar{\mathbf{v}} = [-1,3,0]^T$, respectively. The simulation incorporates $7$ satellites employing the legacy GPS L1 C/A, with their positions and velocities determined from the ephemeris data. Without loss of generality, the satellite clock biases are assumed to be compensated, while the unknown receiver clock bias is set to be $\bar{\delta t} = 10$ ns.

\subsection{Performance Bounds Evaluation}
To evaluate the tightness of the derived performance bounds, we conduct empirical ML estimation under the correctly specified channel models with perfect and statistical channel knowledge, as well as under the misspecified observation model. Fig.~\ref{figure_1} illustrates the $95\%$ horizontal error ellipses obtained from the derived CRBs and MCRB, together with those constructed from the corresponding empirical ML estimation errors. The empirical ellipses are constructed from $N_\mathrm{mc}=1000$ independent Monte Carlo trials to maintain statistical reliability. It can be observed that under both perfect and statistical channel knowledge, the empirical error ellipses closely align with the corresponding CRB ellipses in terms of both size and orientation. The zoomed inset further shows that the difference between the modified CRB and the channel-averaged conditional CRB is numerically negligible, although the former bound is theoretically optimistic. Likewise, the error ellipse of the misspecified ML estimator exhibits a more pronounced elongation and remains in close agreement with the corresponding MCRB ellipse. Collectively, these results demonstrate that the derived performance bounds provide tight characterizations of the empirical estimation error covariance while accurately capturing its directional structure, as reflected by the ellipse elongation and principal-axis orientation.

\subsection{Impact of Receiver Architecture}
\begin{figure}[t]
    \centering
    \includegraphics[width=0.7078069\linewidth]{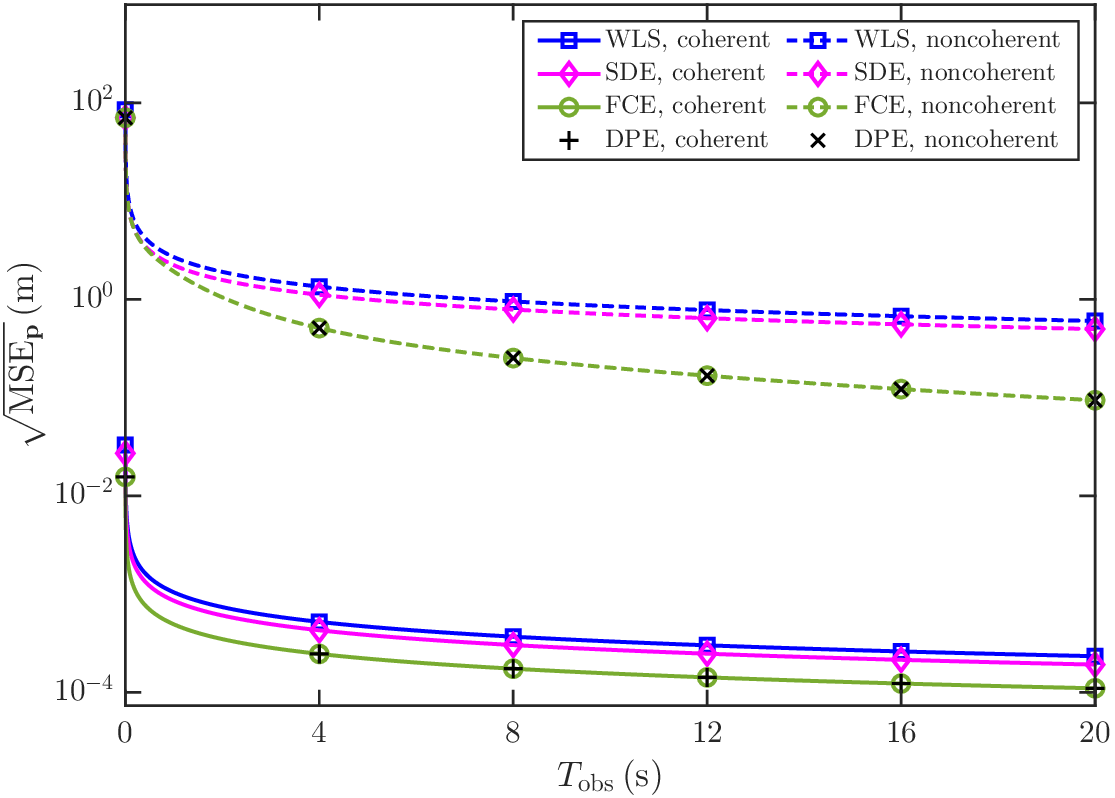}
    \caption{Positioning RMSE versus observation duration for different receiver architectures. Coherent and noncoherent represent the cases where carrier phase information is retained and unavailable for positioning, respectively.}
    \label{figure_2}
\end{figure}

Fig.~\ref{figure_2} studies the impact of receiver architecture on positioning performance under the assumption that the first-stage synchronization-parameter estimates are asymptotically efficient. For comparison, the received $C/N_0$ is fixed for all participating SVs and the multipath-induced uncertainty is excluded from this experiment. Four estimators are examined for both coherent and noncoherent receivers. In particular, DPE represents the one-step estimation benchmark. SDE and FCE employ Fisher-optimal weighting but differ in their second-stage architectures, whereas WLS adopts the same sequentially decoupled two-step architecture as SDE but uses elevation-based weighting.
It can be observed that FCE achieves nearly identical positioning performance to DPE, corroborating their asymptotic equivalence as established in \textbf{Remark~\ref{remark_2}}. In contrast, SDE consistently exhibits a larger root mean square error (RMSE) despite employing Fisher-optimal weighting. This result is consistent with \textbf{Proposition~\ref{proposition_2}}, which indicates that the exclusion of position-related information contained in the Doppler estimates inevitably leads to a performance loss. Moreover, WLS incurs an additional performance loss relative to SDE due to its elevation-based rather than Fisher-optimal weighting. However, this additional loss is substantially smaller than the performance gap between SDE and FCE/DPE, indicating that sequential decoupling has a greater impact on positioning accuracy. It should be noted that the performance gap between SDE and FCE/DPE increases with the observation duration. This trend is more pronounced for noncoherent receivers, where the carrier phase information cannot be exploited for positioning and the relative contribution of Doppler-related position information becomes more significant. 
This is because when carrier phase information is exploitable, the position-related information contains a dominant component on the order of $f_c^2$, which substantially exceeds that contributed by the code envelope scales with $B_\mathrm{eq}^2/12$. As a result, the additional position information carried by the Doppler measurements represents only a small fraction of the total information available for positioning. 
\begin{figure}[t]
    \centering
    \includegraphics[width=0.7078069\linewidth]{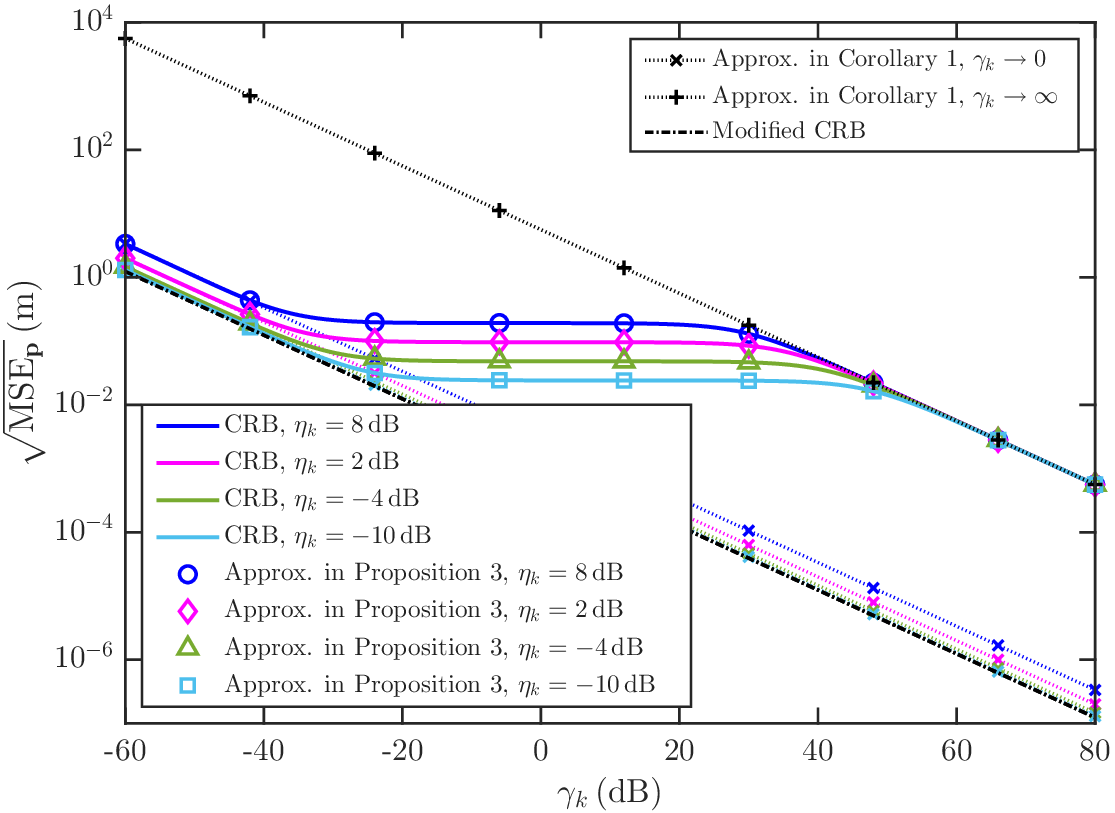}
    \caption{Positioning CRBs versus $\gamma_k$. The bounds are evaluated at different levels of channel randomness.}
    \label{figure_3}
\end{figure}

\subsection{Impact of Channel Uncertainty}
Fig.~\ref{figure_3} examines the impact of $\gamma_k$ on positioning CRBs under the conditions of different fixed levels of channel randomness. To facilitate the characterization of channel randomness, we define $\eta_k\triangleq\chi_{k,2}/\chi_{k,1}$, where a larger $\eta_k$ corresponds to a higher level of channel randomness. For numerical illustration, the same values of $\gamma_k$ and $\eta_k$ are assumed for all participating SVs. It can be observed that the CRBs obtained using the closed-form approximations in \textbf{Proposition~\ref{proposition_4}} are highly consistent with those obtained from the exact FIM over the entire range of $\gamma_k$. Moreover, the CRBs exhibit a clear transition between the low- and high-$\gamma_k$ regimes and approach the corresponding asymptotic expressions in \textbf{Corollary~\ref{corollary_1}} at both extremes. This behavior is consistent with our discussion in \textbf{Remark~\ref{remark_3}} that the information contribution associated with $\mathbf{M}_2$ becomes bounded by channel uncertainty as $\gamma_k$ increases, whereas that associated with $\mathbf{M}_1$ continues to increase.
\begin{figure}[t]
    \centering
    \includegraphics[width=0.7078069\linewidth]{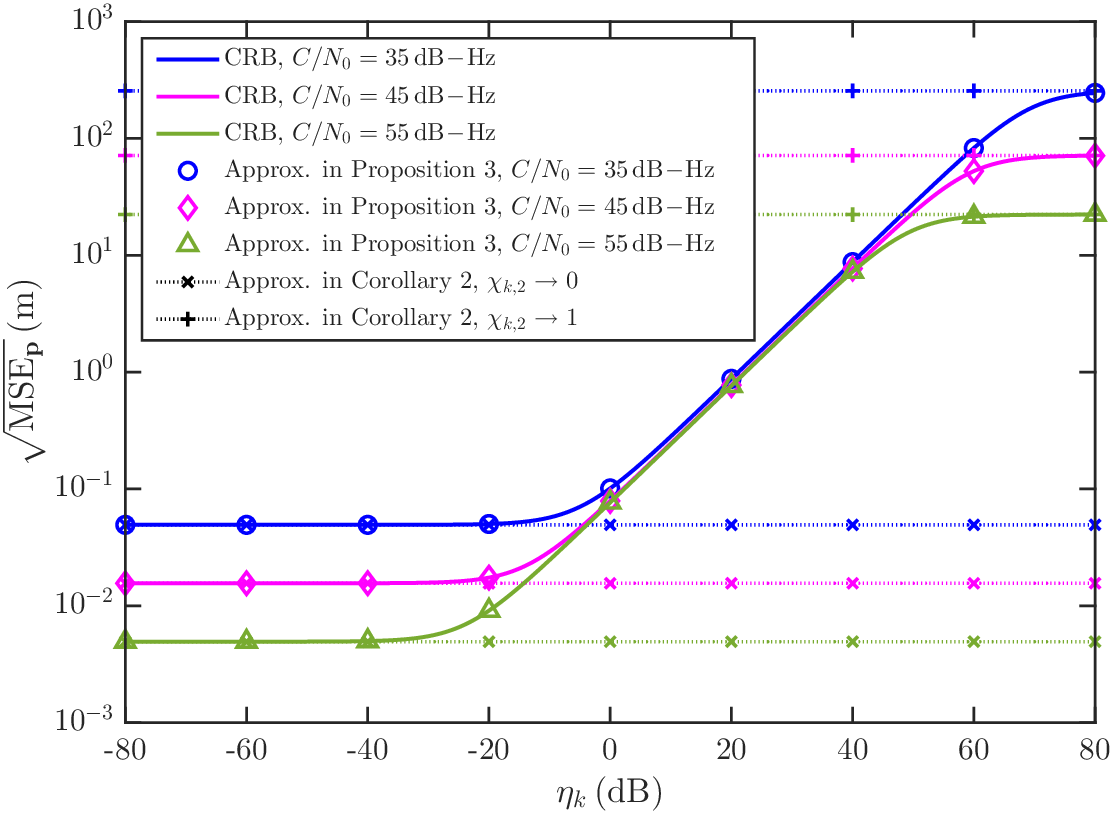}
    \caption{Positioning CRBs versus $\eta_k$. The bounds are evaluated at different levels of received signal quality.}
    \label{figure_4}
\end{figure}
\begin{figure}[t]
    \centering
    \includegraphics[width=0.7078069\linewidth]{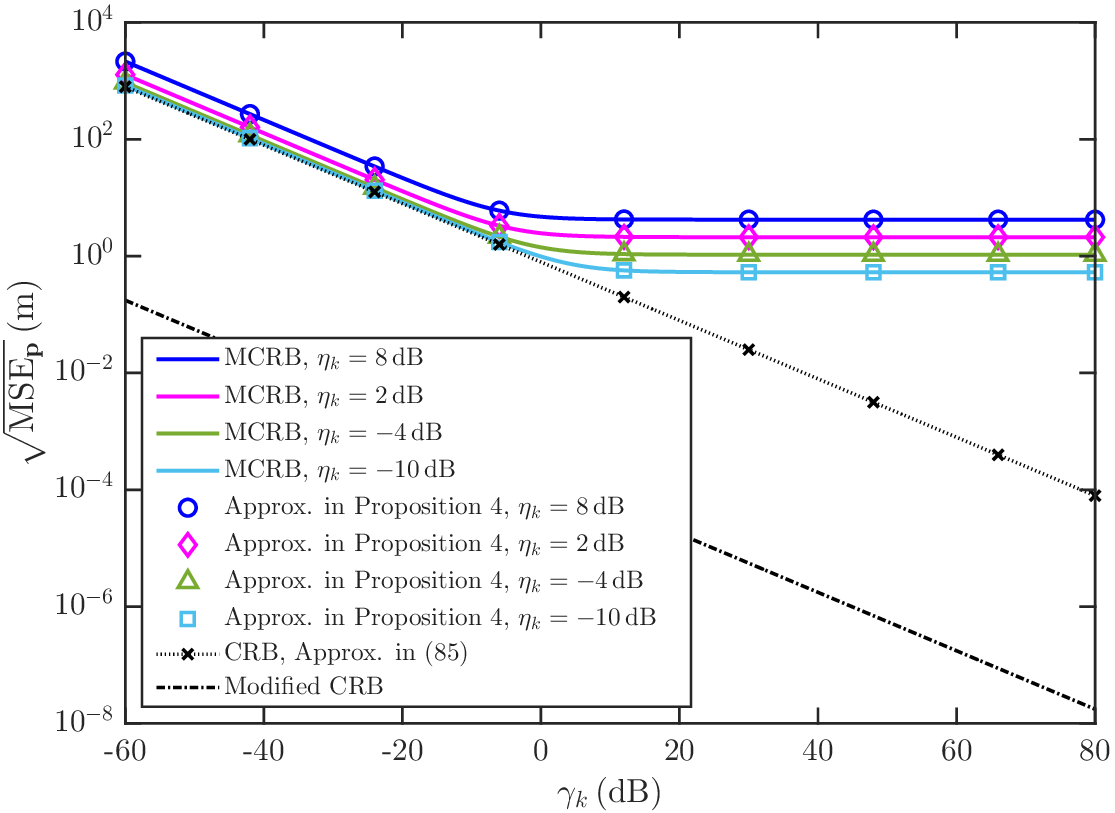}
    \caption{Positioning MCRBs versus $\gamma_k$. The bounds are evaluated at different levels of channel randomness.}
    \label{figure_5}
\end{figure}
\begin{figure}[t]
    \centering
    \includegraphics[width=0.7078069\linewidth]{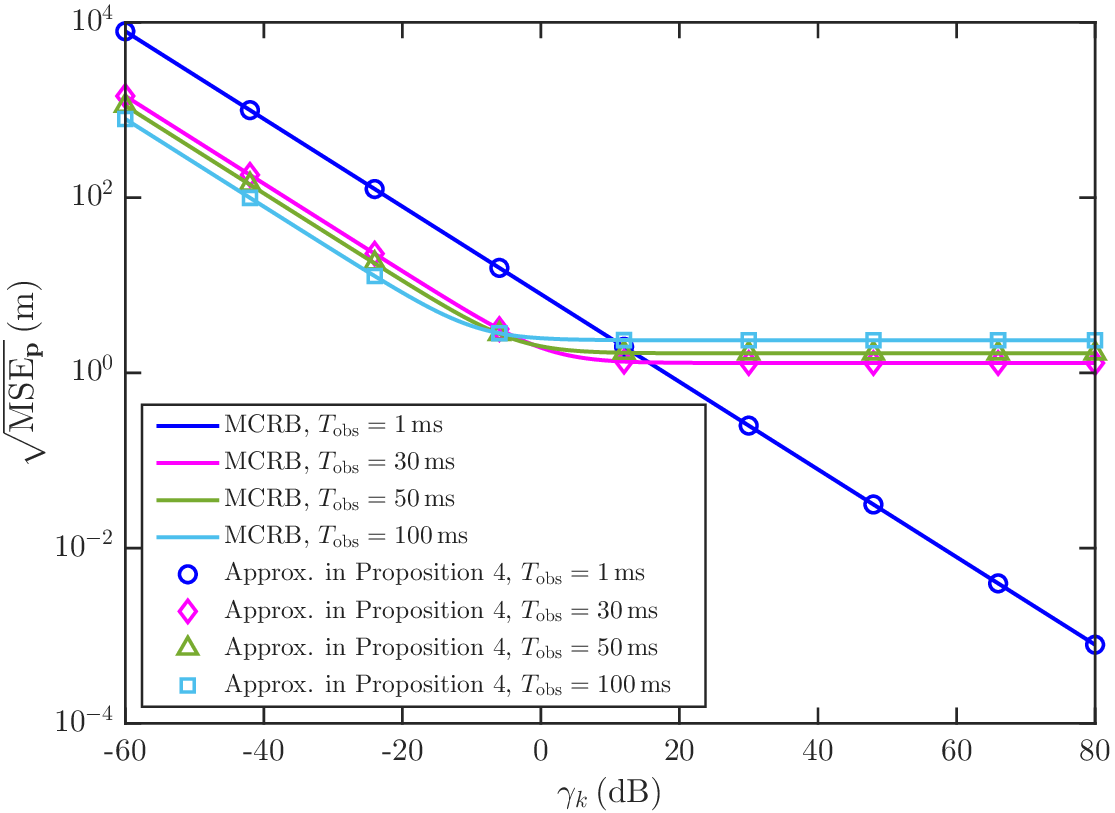}
    \caption{Positioning MCRBs versus $\gamma_k$ with a fixed $\eta_k$. The bounds are evaluated for different observation durations.}
    \label{figure_6}
\end{figure}

Fig.~\ref{figure_4} further investigates the impact of channel randomness on positioning CRBs under the conditions of different fixed levels of $C/N_0$. It can be observed that the CRBs increase with $\eta_k$ and gradually converge to limits that are numerically consistent with the asymptotic expressions in \textbf{Corollary~\ref{corollary_2}}. This result further indicates that the positioning information remains available through the observation covariance even when the coherent mean completely vanishes, as discussed in \textbf{Remark~\ref{remark_4}}. It is interesting to observe that the CRBs corresponding to different levels of $C/N_0$ become nearly identical over the intermediate range of $\eta_k$ from approximately $0$ dB to $40$ dB, indicating that under these multipath conditions, increasing the received signal quality does not necessarily lead to a noticeable improvement in positioning accuracy. This behavior can be explained by the information contribution associated with $\mathbf{M}_2$, which becomes approximately proportional to $1/\eta_k$ and thus largely independent of $\gamma_k$ when $\rho_k\gg1$. The near overlap of the CRBs over this region further suggests that the additional information associated with $\mathbf{M}_1$ provides limited improvement in positioning accuracy.

\subsection{Impact of Model Misspecification}
Fig.~\ref{figure_5} investigates the impact of $\gamma_k$ on positioning MCRBs under the conditions of different fixed levels of channel randomness. It can be observed that the MCRBs obtained using the closed-form approximations of $\mathbf{C}_{\boldsymbol{\xi}}^e$ and $\mathbf{D}_{\boldsymbol{\xi}}^e$ derived from \textbf{Proposition~\ref{proposition_5}} closely agree with those obtained from the exact expressions over the entire $\gamma_k$. As $\gamma_k$ increases beyond approximately $-10$ dB, the MCRBs approach positioning error floors that cannot be eliminated solely by improving the received signal quality due to channel model misspecification. Moreover, we observe that there is an evident gap between the modified CRB and the CRB characterized in~\eqref{sec5-eq23}. This gap reflects the Fisher information loss characterized by $\mathbf{M}-\mathbf{M}_3$ when the complex channel coefficients are treated as unknown nuisance parameters, as discussed in \textbf{Remark~\ref{remark_5}}.
\begin{figure}[t]
    \centering
    \includegraphics[width=0.7078069\linewidth]{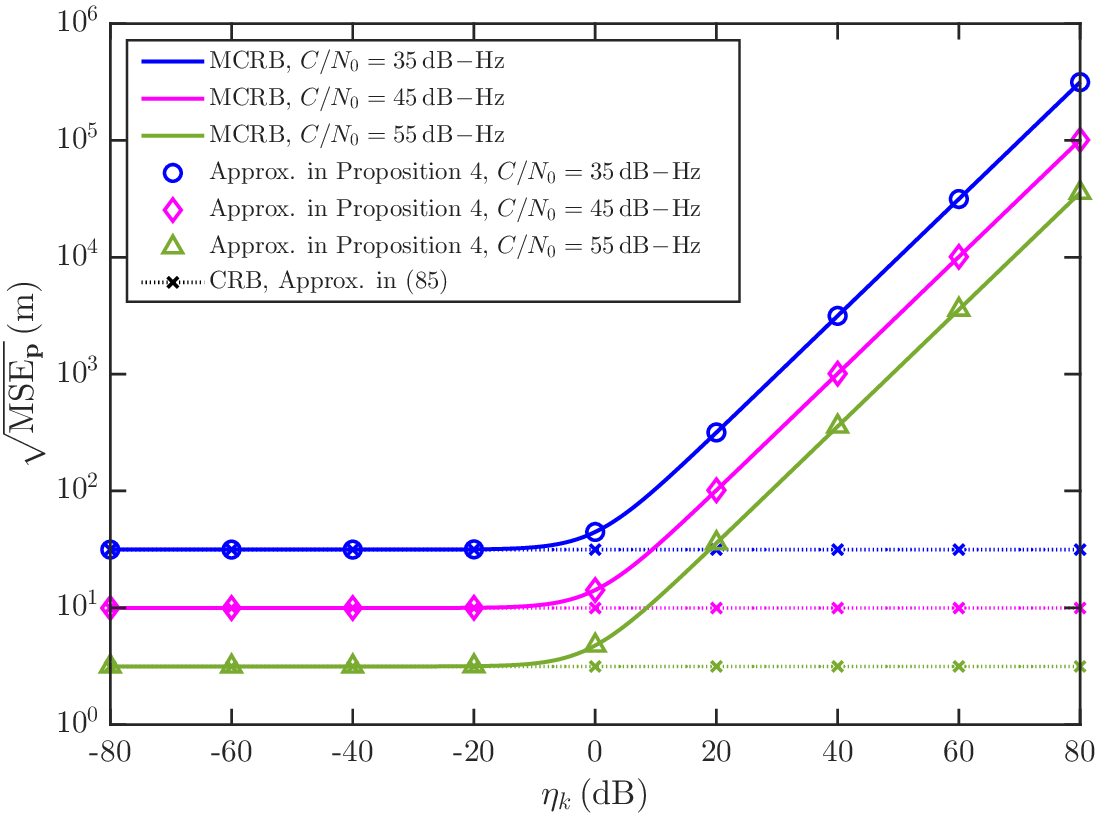}
    \caption{Positioning MCRBs versus $\eta_k$. The bounds are evaluated at different levels of received signal quality.}
    \label{figure_7}
\end{figure}
Fig.~\ref{figure_6} further examines the impact of channel model misspecification on positioning MCRBs under different observation durations. It can be observed that increasing $T_{\mathrm{obs}}$ improves positioning accuracy in the low-$\gamma_k$ regime. However, as $\gamma_k$ increases, the MCRBs approach positioning error floors that become more pronounced with longer observation durations. It can be explained by the additional score covariance $\Delta\mathbf{D}_{\boldsymbol{\xi}}^e$, which depends on the observation duration through $\left(T_\mathrm{obs}^2-T_\mathrm{code}^2\right)/12$.
When $T_{\mathrm{obs}}=T_{\mathrm{code}}$, this term vanishes and the MCRB reduces to the conventional CRB, which continues to decrease with $\gamma_k$ without exhibiting an error floor, as illustrated by the blue curve. For longer observation durations, the unmodeled random channel variations across code periods introduce an additional contribution to the MCRB, giving rise to the observed positioning error floors. Consequently, the resulting performance degradation may outweigh the benefits of extended observations, indicating that increasing the observation duration does not necessarily improve positioning accuracy under channel model misspecification.

Fig.~\ref{figure_7} studies the impact of channel randomness on positioning MCRBs under the conditions of different fixed levels of $C/N_0$. It can be observed that as $\eta_k\rightarrow0$, the MCRBs approach the CRBs approximated in~\eqref{sec5-eq23}. It is because the assumed channel model becomes correctly specified in the absence of channel randomness. However, as $\eta_k$ increases, the MCRBs increase significantly and eventually diverge as the coherent mean approaches zero. In contrast to the finite limits of the CRBs observed in Fig.~\ref{figure_4}, this result further demonstrates that the positioning information carried by the observation covariance cannot be effectively exploited when the receiver neglects the random channel variations.

\section{Conclusion}\label{section_7}
This paper investigated Fisher information availability and utilization in GNSS DPE under channel uncertainty induced by short-delay multipath. A unified estimation-theoretic framework demonstrated the asymptotic equivalence between DPE and fully coupled two-step estimation, while identifying the performance loss caused by sequential decoupling even with Fisher-optimal weighting. Leveraging a scalar-equivalent statistical channel model, receiver-state CRBs and MCRBs were derived to characterize the effects of channel knowledge and model misspecification. Our results demonstrated that under statistical channel knowledge, channel randomness limits certain components of waveform-domain Fisher information, thereby reducing the positioning benefits of improved received signal quality in some regimes. Nevertheless, positioning information remains available through the observation covariance even when the coherent channel mean vanishes. In contrast, neglecting channel randomness under the considered misspecified model can introduce positioning error floors and lead to performance degradation as the coherent mean approaches zero. Collectively, these findings clarify how channel uncertainty affects the Fisher information available for positioning, how receiver architecture determines its utilization, and how model misspecification impacts the estimation performance.

\appendices
\section{Proof of Proposition~\ref{proposition_2}}\label{appendix_2}
To verify the positive semidefiniteness of the covariance difference, for any real vector $\mathbf{u}\neq\mathbf{0}$, define
\begin{align}
    \mathbf{a} &= \mathbf{F}^{-1/2}(\boldsymbol{\zeta}) \mathbf{E}_{\mathrm{dec}}^T \mathbf{u},\label{appendixA-eq1}\\
    \mathbf{b} &= \mathbf{F}^{1/2}(\boldsymbol{\zeta})\mathbf{G}\left(\mathbf{G}^T\mathbf{F}(\boldsymbol{\zeta})\mathbf{G}\right)^{-1}\mathbf{u}. \label{appendixA-eq2}
\end{align}
Using the Cauchy-Schwarz inequality, we have
\begin{equation}\label{appendixA-eq3}
    \begin{aligned}
        &\Vert\mathbf{a}\Vert^2\Vert\mathbf{b}\Vert^2 - \left(\mathbf{a}^T\mathbf{b}\right)^2 =\left(
        \mathbf{u}^T\mathbf{E}_{\mathrm{dec}}\mathbf{F}^{-1}(\boldsymbol{\zeta})\mathbf{E}_{\mathrm{dec}}^T\mathbf{u}
        \right)\\
        &\times\left(
        \mathbf{u}^T\left(\mathbf{G}^T\mathbf{F}(\boldsymbol{\zeta})\mathbf{G}\right)^{-1}\mathbf{u}
        \right) -
        \left(
        \mathbf{u}^T\left(\mathbf{G}^T\mathbf{F}(\boldsymbol{\zeta})\mathbf{G}\right)^{-1}\mathbf{u}
        \right)^2 \geq 0.
    \end{aligned}
\end{equation}
Since $\mathbf{u}^T \left(\mathbf{G}^T\mathbf{F}(\boldsymbol{\zeta})\mathbf{G}\right)^{-1}
\mathbf{u}>0$, dividing~\eqref{appendixA-eq3} by this term and using~\eqref{sec3-eq12} yields
\begin{equation}\label{appendixA-eq4}
    \mathbf{u}^T\left(
    \mathbf{E}_{\mathrm{dec}}\mathbf{F}^{-1}(\boldsymbol{\zeta})\mathbf{E}_{\mathrm{dec}}^T- \mathbf{E}_{\mathrm{opt}}
    \mathbf{F}^{-1}(\boldsymbol{\zeta})\mathbf{E}_{\mathrm{opt}}^T
    \right)\mathbf{u} \geq 0, 
\end{equation}
which yields
\begin{equation}\label{appendixA-eq5}
    \mathbf{E}_{\mathrm{dec}}\mathbf{F}^{-1}(\boldsymbol{\zeta})\mathbf{E}_{\mathrm{dec}}^T - \mathbf{E}_{\mathrm{opt}}
    \mathbf{F}^{-1}(\boldsymbol{\zeta})\mathbf{E}_{\mathrm{opt}}^T = \Delta\mathbf{E}\mathbf{F}^{-1}(\boldsymbol{\zeta})\Delta\mathbf{E}^T
    \succeq\mathbf{0}.
\end{equation}
The proof is thus completed.
\section{Proof of Proposition~\ref{proposition_3}}\label{appendix_3}
For notational simplicity, we use $\dot{(\cdot)}^\tau$ and  $\dot{(\cdot)}^\nu$ to define the derivatives with respect to $\tau_k$ and $\nu_k$ respectively. The partial derivatives of $s_{p,n,k}$ with respect to $\tau_k$ and $\nu_k$ are given by
\begin{align}
    \dot{s}_{p,n,k}^\tau
    &= e^{\jmath\psi_{p,n,k}}\left(
    \dot{x}_{n,k}^{\tau}-\jmath2\pi f_cx_{n,k}
    \right),\label{appendixB-eq1}\\
    \dot{s}_{p,n,k}^\nu 
    &= \jmath2\pi t_{p,n}x_{n,k}e^{\jmath\psi_{p,n,k}},\label{appendixB-eq2}
\end{align}
where $\psi_{p,n,k}\triangleq2\pi(\nu_kt_{p,n}-f_c\tau_k)$, $t_{p,n}\triangleq pT_\mathrm{code} + nT_\mathrm{samp}$. For the real-valued code envelope, the cross terms involving its delay derivative have zero real part. Therefore, the Gram matrix can be calculated as
\begin{equation}\label{appendixB-eq3}
    \Re\left\{\mathbf{J}_k^H\mathbf{J}_k\right\} = 4\pi^2E_s\begin{bmatrix}
        P\left(
        f_c^2+\frac{\Vert\dot{\mathbf{x}}_k^{\tau}\Vert^2}{4\pi^2E_s}
        \right) & -f_c\sum_{p}\bar{t}_{p,k}\\
        -f_c\sum_{p}\bar{t}_{p,k} & \sum_{p}\bar{t^2}_{p,k}
    \end{bmatrix},
\end{equation}
where $\bar{t}_{p,k}\triangleq\frac{1}{E_s}\sum_n t_{p,n}x^2_{p,n}$, $\bar{t^2}_{p,k}\triangleq\frac{1}{E_s}\sum_n t^2_{p,n}x^2_{p,n}$, and $\Vert\dot{\mathbf{x}}_k^{\tau}\Vert^2 = 4\pi^2\int f^2\vert X_k(f)\vert^2\,df$. Considering uniform power spectral density for the signal of interest~\cite{sec1-ref10}, we have $\Vert\dot{\mathbf{x}}^\tau_k\Vert^2 \simeq \pi^2E_sB^2_{\mathrm{eq}}/3$. Considering uniform temporal energy over the observation, we have
\begin{align}
    &\sum_{p}\bar{t}_{p,k} \simeq PT_\mathrm{obs}/2,&
    \sum_{p}\bar{t^2}_{p,k} \simeq PT_\mathrm{obs}^2/3.\label{appendixB-eq6}
\end{align}
Therefore,~\eqref{appendixB-eq3} can be rewritten as
\begin{equation}\label{appendixB-eq7}
    \Re\{\mathbf{J}_k^H\mathbf{J}_k\} \simeq 4\pi^2PE_s\begin{bmatrix}
        B_{\mathrm{eq}}^2/12 +  f_c^2& -f_cT_{\mathrm{obs}}/2\\
        -f_cT_{\mathrm{obs}}/2 & T_{\mathrm{obs}}^2/3
    \end{bmatrix}.
\end{equation}
Substituting~\eqref{appendixB-eq7} into~\eqref{sec4-eq3} yields~\eqref{sec4-eq4}. The proof is thus completed.
\section{Proof of Proposition~\ref{proposition_4}}\label{appendix_4}
Using~\eqref{appendixB-eq1} and~\eqref{appendixB-eq2}, the Gram matrix of the waveform Jacobian projected onto $\mathcal{S}_k$ can be obtained as
\begin{equation}\label{appendixC-eq1}
    \Re\left\{
    \mathbf{J}_k^H\mathbf{P}_k\mathbf{J}_k
    \right\} = 4\pi^2E_s\begin{bmatrix}
        Pf_c^2 & -f_c\sum_{p}\bar{t}_{p,k}\\[0.5 em]
        -f_c\sum_{p}\bar{t}_{p,k} & \sum_{p}\bar{t}_{p,k}^2
    \end{bmatrix}.
\end{equation}
The uniform temporal energy assumption yields $\sum_p\bar{t}_{p,k}^2\simeq P\left(T_\mathrm{obs}^2/3 - T_\mathrm{code}^2/12\right)$.
Therefore,~\eqref{appendixC-eq1} can be rewritten as
\begin{equation}\label{appendixC-eq4}
    \Re\{\mathbf{J}_k^H\mathbf{P}_k\mathbf{J}_k\} \simeq4\pi^2PE_s\begin{bmatrix}
            f_c^2 & -f_cT_{\mathrm{obs}}/2\\
            -f_cT_{\mathrm{obs}}/2 & T_{\mathrm{obs}}^2/3 - T_{\mathrm{code}}^2/12
        \end{bmatrix}.
\end{equation}
Furthermore, using
$\mathbf{P}_k^\perp=\mathbf{I}_{PN}-\mathbf{P}_k$
and~\eqref{appendixB-eq7} gives
\begin{equation}\label{appendixC-eq5}
    \Re\{\mathbf{J}_k^H\mathbf{P}_k^\perp\mathbf{J}_k\}
\simeq4\pi^2PE_s\begin{bmatrix}
            B_{\mathrm{eq}}^2/12 & 0\\
            0 & T_{\mathrm{code}}^2/12
        \end{bmatrix}.
\end{equation}
Substituting~\eqref{appendixC-eq4} and~\eqref{appendixC-eq5} into~\eqref{sec4-eq12} and~\eqref{sec4-eq15}
yields~\eqref{sec4-eq16}. The proof is thus completed.

\section{Proof of Proposition~\ref{proposition_5}}\label{appendix_5}
Define $\mathbf{Q}_k\triangleq\frac{1}{PE_s} \mathbf{s}_k\mathbf{s}_k^H$ as the orthogonal projection matrix onto the subspace spanned by $\mathbf{s}_k$, and $\mathbf{Q}_k^\perp\triangleq\mathbf{I}_{PN} -\mathbf{Q}_k$ as the corresponding projector onto its orthogonal complement. Then, using~\eqref{sec5-eq11} and ~\eqref{sec5-eq12}, we obtain
\begin{equation}\label{appendixD-eq1}
\mathbf{C}_{k,\boldsymbol{\zeta}\mathbf{a}}\mathbf{C}_{k,\mathbf{a}\mathbf{a}}^{-1}\mathbf{C}_{k,\boldsymbol{\zeta}\mathbf{a}}^T 
 = \frac{2\vert\mu_{\tilde{h},k}\vert^2}{\sigma_{z,k}^2}\Re\left\{\mathbf{J}_k^H\mathbf{Q}_k\mathbf{J}_k\right\}.
\end{equation}
Using equations~\eqref{appendixB-eq1},~\eqref{appendixB-eq2}, together with the corresponding closed-form results derived under the temporal energy assumption, we have $\mathbf{s}_k^H\mathbf{J}_k = \jmath2\pi PE_s\begin{bmatrix}
    -f_c & T_\mathrm{obs}/2
\end{bmatrix}$. Then, the Gram matrix in~\eqref{appendixD-eq1} can be expressed as
\begin{equation}\label{appendixD-eq2}
    \Re\left\{\mathbf{J}_k^H\mathbf{Q}_k\mathbf{J}_k\right\}\simeq4\pi^2PE_s\begin{bmatrix}
        f_c^2 & -f_cT_\mathrm{obs}/2\\
        -f_cT_\mathrm{obs}/2\ & T_\mathrm{obs}^2/4
    \end{bmatrix}.
\end{equation}
Therefore, it follows from~\eqref{sec5-eq13} that the equivalent negative expected Hessian can be calculated as 
\begin{equation}\label{appendixD-eq3}
\begin{aligned}
    \mathbf{C}_k^e &= \frac{2\vert\mu_{\tilde{h},k}\vert^2}{\sigma_{z,k}^2}\Re\left\{\mathbf{J}_k^H\mathbf{Q}_k^\perp\mathbf{J}_k\right\}\\
    &\simeq8\pi^2PE_s\gamma_k\chi_{k,1}\begin{bmatrix}
            B_{\rm eq}^2/12 & 0\\
            0 & T_{\rm obs}^2/12
        \end{bmatrix}.
\end{aligned}
\end{equation}
To derive $\mathbf{D}_k^e$, we first eliminate the nuisance complex gain from the score function. The resulting equivalent score evaluated at the pseudo-true parameters is given by $\mathbf{g}_{\boldsymbol{\zeta},k}^e = \frac{2}{\sigma_{z,k}^2}
\Re\left\{
\mu_{\tilde h,k}^*\mathbf J_k^H
\mathbf Q_k^\perp
(\mathbf y_k-\mu_{\tilde h,k}\mathbf s_k)
\right\}$. Since the residual follows a proper complex Gaussian distribution with covariance $\mathbf\Sigma_k$, the equivalent score covariance is obtained as
\begin{equation}\label{appendixD-eq4}
    \mathbf{D}_k^e =
    \frac{2|\mu_{\tilde h,k}|^2}{\sigma_{z,k}^4}
    \Re\left\{
    \mathbf J_k^H\mathbf Q_k^\perp
    \mathbf\Sigma_k\mathbf Q_k^\perp\mathbf J_k
    \right\}. 
\end{equation}
Since $\mathbf{s}_k = \mathbf{S}_k\mathbf{1}_P$, we have $\mathrm{span}(\mathbf{s}_k)\subseteq\mathrm{col}(\mathbf{S}_k)$, which implies $\mathbf{P}_k\mathbf{Q}_k =\mathbf{Q}_k\mathbf{P}_k=\mathbf{Q}_k$. Since $\mathbf{\Sigma}_k = \sigma_{z,k}^2(\mathbf{I}_{PN} + \rho_k\mathbf{P}_k)$ and $\left(\mathbf{Q}_k^\perp\right)^2 = \mathbf{Q}_k^\perp$, we obtain
\begin{equation}\label{appendixD-eq5}
    \mathbf{Q}_k^\perp\mathbf{\Sigma}_k
    \mathbf{Q}_k^\perp
    =\sigma_{z,k}^2\left[
    \mathbf{Q}_k^\perp + \rho_k(\mathbf{P}_k-\mathbf{Q}_k)\right].
\end{equation}
Furthermore, since $\mathbf{P}_k-\mathbf{Q}_k =\mathbf{Q}_k^\perp-\mathbf{P}_k^\perp$, we have
\begin{equation}\label{appendixD-eq6}
    \Re\{\mathbf{J}_k^H(\mathbf{P}_k-\mathbf{Q}_k) \mathbf{J}_k\}
     =\Re\left\{\mathbf{J}_k^H\mathbf{Q}_k^\perp\mathbf{J}_k\right\} - \Re\left\{\mathbf{J}_k^H\mathbf{P}_k^\perp\mathbf{J}_k\right\}
     .
\end{equation}
Substituting~\eqref{appendixD-eq5} into~\eqref{appendixD-eq4} and using~\eqref{appendixD-eq3} and~\eqref{appendixC-eq5} to evaluate~\eqref{appendixD-eq6}, we obtain~\eqref{sec5-eq16}.
The proof is thus completed.

\vfill

\end{document}